%% file: taggpt.tex
\documentclass{article}
\pdfoutput=1

\usepackage[preprint]{neurips_2022}

\usepackage[utf8]{inputenc} 
\usepackage[T1]{fontenc}    
\usepackage{hyperref}       
\usepackage{url}            
\usepackage{booktabs}       
\usepackage{amsfonts}       
\usepackage{nicefrac}       
\usepackage{microtype}      
\usepackage{xcolor}         
\usepackage{graphicx}
\usepackage{pifont}
\usepackage{CJKutf8}
\usepackage{makecell}
\usepackage{multirow}
\usepackage{xspace}
\usepackage{subfigure}
\usepackage{longtable}
\usepackage{float}
\setcitestyle{numbers,square}

\title{TagGPT: Large Language Models are Zero-shot Multimodal Taggers}

\def\modelname{\texttt{TagGPT}\xspace}
\author{%
 Chen Li$^1$ \qquad Yixiao Ge$^1$ \qquad Jiayong Mao$^2$ \qquad Dian Li$^2$ \qquad Ying Shan$^1$ \\~\\
 $^1$ARC Lab, Tencent PCG \\
 $^2$Foundation Technology Center, Tencent PCG \\
}

\begin{document}

\maketitle

\begin{abstract}
Tags are pivotal in facilitating the effective distribution of multimedia content in various applications in the contemporary Internet era, such as search engines and recommendation systems. 
%
%
%
Recently, large language models (LLMs) have demonstrated impressive capabilities across a wide range of tasks.
In this work, we propose \modelname, a fully automated system capable of tag extraction and multimodal tagging in a completely zero-shot fashion.
Our core insight is that, through elaborate prompt engineering, LLMs are able to extract and reason about proper tags given textual clues of multimodal data, \textit{e.g.}, OCR, ASR, title, \textit{etc}.
Specifically, to automatically build a high-quality tag set that reflects user intent and interests for a specific application, \modelname predicts large-scale candidate tags from a series of raw data via prompting LLMs, filtered with frequency and semantics.
Given a new entity that needs tagging for distribution, \modelname introduces two alternative options for zero-shot tagging, \textit{i.e.}, a generative method with late semantic matching with the tag set, and another selective method with early matching in prompts.
It is well noticed that \modelname provides a system-level solution based on a modular framework equipped with a pre-trained LLM (GPT-3.5 used here) and a sentence embedding model (SimCSE used here), which can be seamlessly replaced with any more advanced one you want.
\modelname is applicable for various modalities of data in modern social media and showcases strong generalization ability to a wide range of applications. We evaluate \modelname on publicly available datasets, \textit{i.e.}, Kuaishou and Food.com, and demonstrate the effectiveness of \modelname compared to existing hashtags and off-the-shelf taggers.\footnote{Project page: \url{https://github.com/TencentARC/TagGPT}}

\end{abstract}

\input{texts/1.introduction}
\input{texts/2.taggpt}
\input{texts/3.experiments}
\input{texts/4.limitations}
\input{texts/5.conclusion}

\subsection*{Acknowledgements}
The project of \modelname\ is still under development with the aim of an industrial-strength zero-shot tagger. \modelname\ is a collaborative work of ARC-QQ Joint Lab at Tencent PCG, with the help of Zekun Wang, Yunxuan Zhang, Kun Yi, Shupeng Su, and Shansong Liu. 

\input{texts/6.appendix}

\bibliographystyle{plain}
\bibliography{taggpt}

\end{document}

%% file: texts/1.introduction.tex
\section{Introduction}\label{sec:intro}

\begin{figure}[ht]
    \centering
    \includegraphics[width=.95\linewidth]{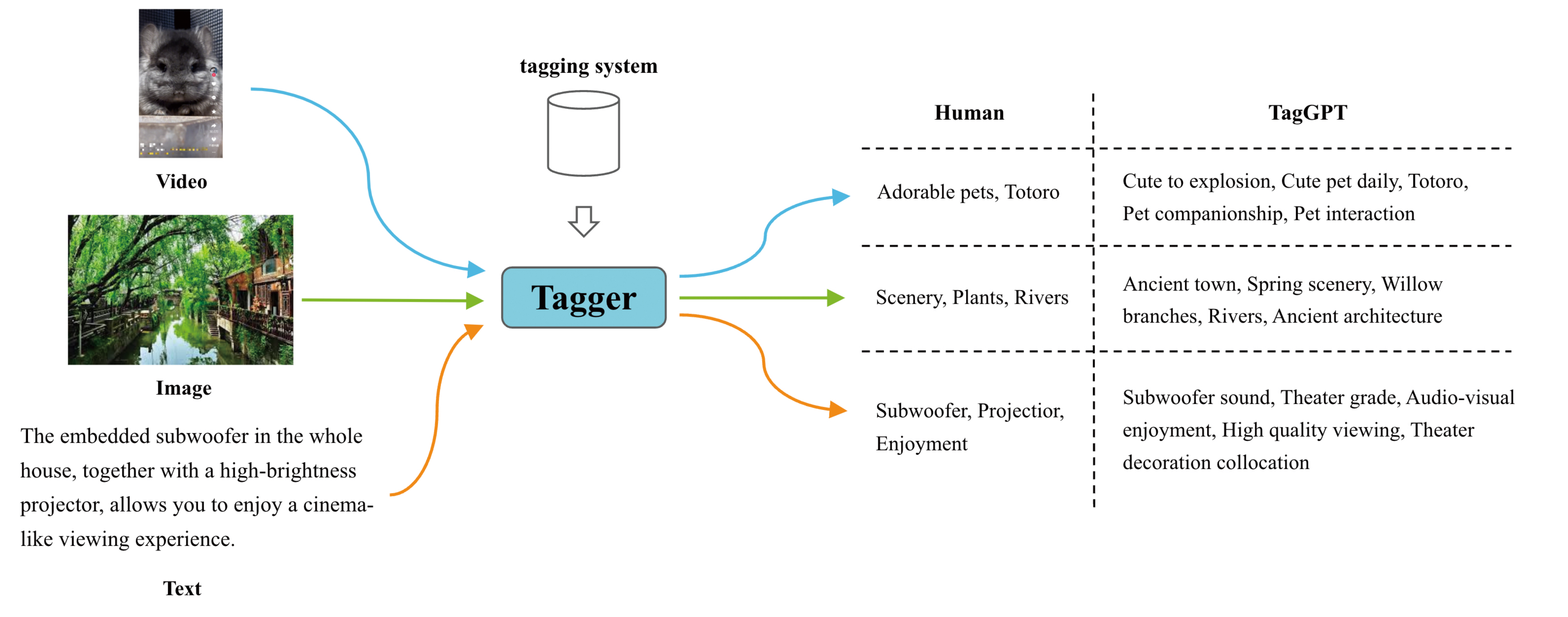}
    \caption{Given multimodal content from social media (\textit{e.g.}, Twitter, Weibo, \textit{etc.}), a tagger aims to produce several phrases that can properly describe the content and reflect the user's interests. }
    \label{fig:example}
\end{figure}

Tags, as concise descriptions of semantic content, have been demonstrated to play an incredibly important role in numerous downstream tasks~\cite{tag_survey2010kdd}. They facilitate the system's quick and accurate understanding of user interests and search intentions, simplifying the search~\cite{tag_search2016semantics, tag_search_22010influence} and recommendation~\cite{subjective2022aaai, tag_recomm_22022nc, tag_recommm2023icsadl} process as a result. For example, as illustrated in Fig.~\ref{fig:example}, on social media platforms with multimodal content, users are permitted to use customized tags to effortlessly describe their published content, luring users with similar interests. Therefore, obtaining top-notch tags and constructing comprehensive tagging systems and corresponding taggers have always been important ways to improve the service quality of such applications.

The traditional sources for obtaining tags are primarily human annotation and language model extraction. However, tags obtained through manual annotation tend to be subjective and expensive~\cite{subjective2022aaai}, while those extracted by language models are constrained by limited understanding capabilities and may lack reliability~\cite{tag_survey2010kdd}. Especially with the rapid expansion of the semantic space in today's social networks and the increasing diversity of data modalities, the challenge of quickly obtaining high-quality tags from massive multi-modal data has become a pressing research and application issue.

To construct a high-quality content-oriented tagging system and corresponding zero-shot tagger, we propose \modelname, a generative solution in this paper. Our approach draws inspiration from the impressive inference capabilities of Large Language Models (LLMs) and their exceptional performance in zero-shot settings~\cite{gpt32020arxiv,gpt42023arxiv,bloom2023arxiv,llama2023arxiv,palm2022arxiv}. We begin by converting multimodal data into unified textual clues through mature unimodal models. These clues are then inserted into a discrete instruction prompt, which feeds into an LLM to generate potential tags. By collecting tags from large-scale inputs, we can collate a set of content tags. We then utilize post-processing techniques such as unsupervised semantic similarity fusion and frequency distribution filtering to obtain a high-quality content-oriented tagging system. Finally, we design two distinct zero-shot taggers that rely on the generation or context inference abilities of LLMs for zero-shot data tagging.

We have built benchmarks using data from two popular applications and conducted experiments. Our experimental results have shown that \modelname has the ability to create a top-notch tagging system and allocate suitable tags to the input multimodal data in the zero-shot scenario. For the complete project code and additional examples, please refer to \url{https://github.com/TencentARC/TagGPT}.

%% file: texts/2.taggpt.tex
\section{TagGPT}\label{sec:taggpt}
\modelname is a framework designed for processing multimodal content. The framework boasts a pipeline that automatically constructs a tagging system, along with a tagger that completes zero-shot annotation. At the heart of this framework lies the ability to leverage the capability of LLMs to obtain a high-performance, top-quality tagging system, all at a low cost. In the following sections, we will introduce the two primary modules that comprise \modelname.

\subsection{Tagging System Construction}
\begin{figure}[ht]
    \centering
    \includegraphics[width=.95\linewidth]{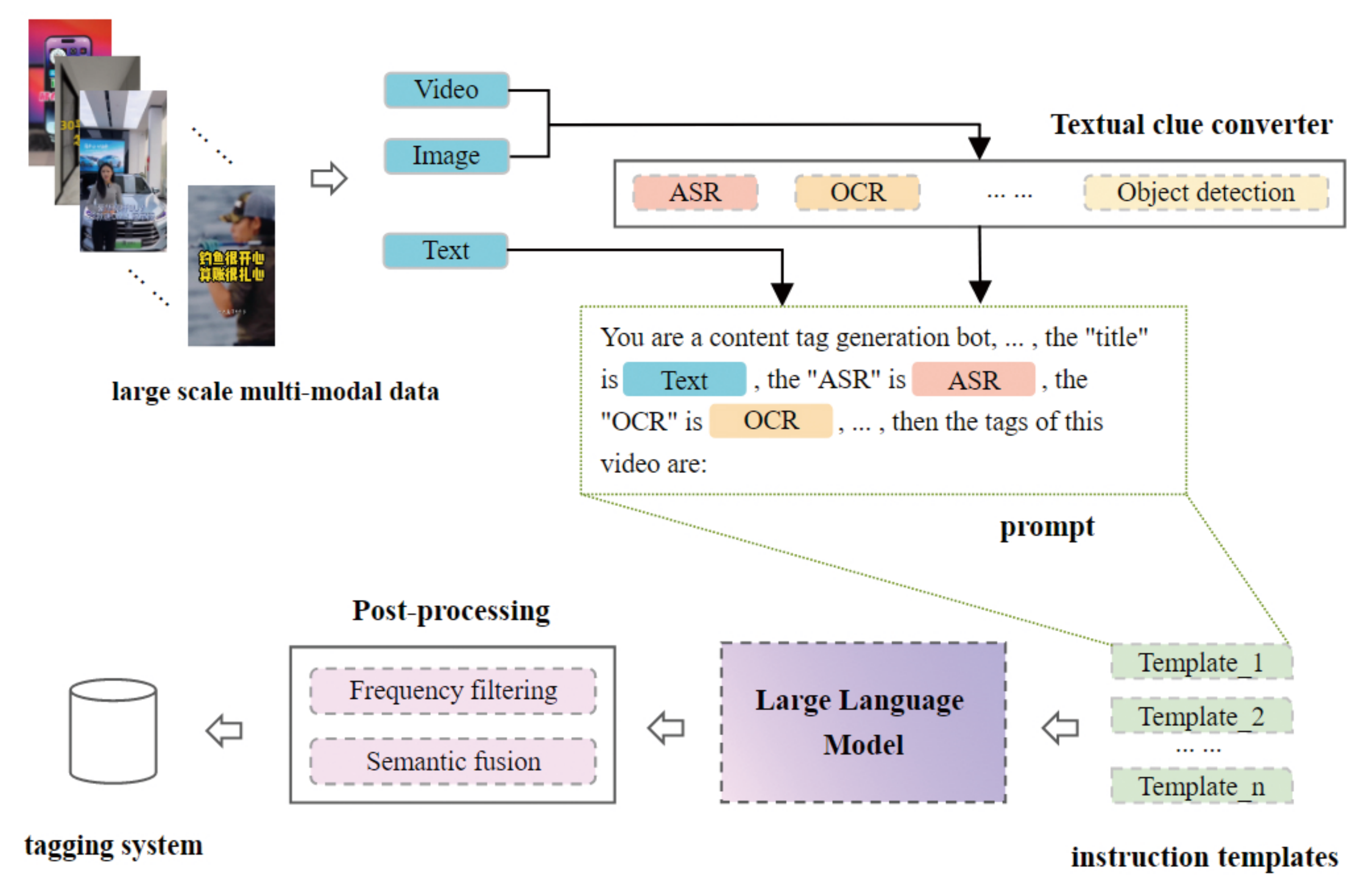}
    \caption{Given a series of raw data from a specific application, \modelname\ is capable of building a high-quality tagging system in an entirely zero-shot manner without extra knowledge or human annotation. Such a paradigm enables instant tagging of new applications with zero labor cost.}
    \label{fig:tag_system}
\end{figure}

The construction of a content-oriented tagging system aims to capture valuable semantic information from large-scale data and form a complete tagging system. In \modelname, to be able to mine high-quality tags from large-scale multimodal data, as shown in Fig.~\ref{fig:tag_system}, we propose a pipeline consisting of three sub-modules.

\subsubsection{Textual clue converter}
With the advancement of mobile networks and multimedia technology, popular applications like social networks, e-commerce platforms, and media information platforms are increasingly utilizing multimodal data as the primary information distribution carrier. The rapid development of a sound tagging system based on such multimodal data can provide fundamental support for services such as search and recommendation. 

Recently, LLM has performed well in content understanding, reasoning, and generation tasks, thus we attempt to apply LLM to the construction of tagging systems. To maximize the benefits of LLMs, \modelname inspired by~\cite{socratic022arxiv,tc2022arxiv} first processes multimodal data and extracts rich textual clues from it with the help of various unimodal basic models as input to downstream modules. For example, when dealing with video data, \modelname employs robust Automatic Speech Recognition (ASR)~\cite{asr2016automatic} and Optical Character Recognition (OCR)~\cite{ocr2017optical} models to extract textual information from video frames and audio. Similarly, to meet the construction requirements of the tagging system in various domains, processing technologies such as object detection~\cite{od2020object} and action recognition~\cite{har2022human} can be utilized to provide textual clues from different perspectives for \modelname.

\subsubsection{Tag generation based on LLM}
With the ever-evolving progress of LLMs in terms of training data scale, parameter scale, and training strategies, LLMs such as GPT~\cite{gpt32020arxiv,gpt42023arxiv} and PaLM~\cite{palm2022arxiv} have exhibited remarkable understanding, reasoning, and generation capabilities. Consequently, numerous downstream tasks have benefited from these models, achieving significant performance improvements. Recently, the introduction of InstructGPT~\cite{instructgpt2022training} has further fortified the ability of LLMs to comprehend various task instructions, thus enhancing their capacity to transition between different tasks. In \modelname, we strive to incorporate the formidable instruction understanding ability of LLMs into new tasks, as well as their potential to understand, reason, and generate content. This will facilitate the creation of a tagging system. More specifically, we have created multiple instruction templates for acquiring tags and filled in the corresponding positions with textual clues that correspond to the single multimodal data provided from upstream. The input text is then fed into the LLM, and the corresponding tag result can be obtained. After a large batch of generation operations, large-scale tags generated by LLMs can be obtained, resulting in a complete and comprehensive candidate tag set.

\subsubsection{Post-processing}
Of course, the tag set constructed in this manner is bound to contain a lot of noise and redundant semantics. Thus, \modelname includes a post-processing module at the end of the tagging system construction in hopes of improving the quality of the tagging system through a series of operations. Firstly, the module truncates tags that are too high or too low based on their frequency; tags with a high frequency are likely to come from popularity deviation and lack sample distinguishability~\cite{zhang2021causal}, while tags with low frequency may come from very few samples and lack significant tag value~\cite{xu2006towards}. The module then checks the semantic similarity of the remaining tags and fuses those with similar semantics to further reduce the tagging system's scale. Specifically, all tags in the tagging system are encoded based on an unsupervised pre-trained text encoder to obtain their corresponding representations. The module then calculates the cosine similarity to determine whether the tags should be fused.

So far, we can automatically construct a corresponding tagging system based on given large-scale multimodal data.

\subsection{Zero-shot Multimodal Tagger}
\begin{figure}[ht]
    \centering
    \includegraphics[width=.95\linewidth]{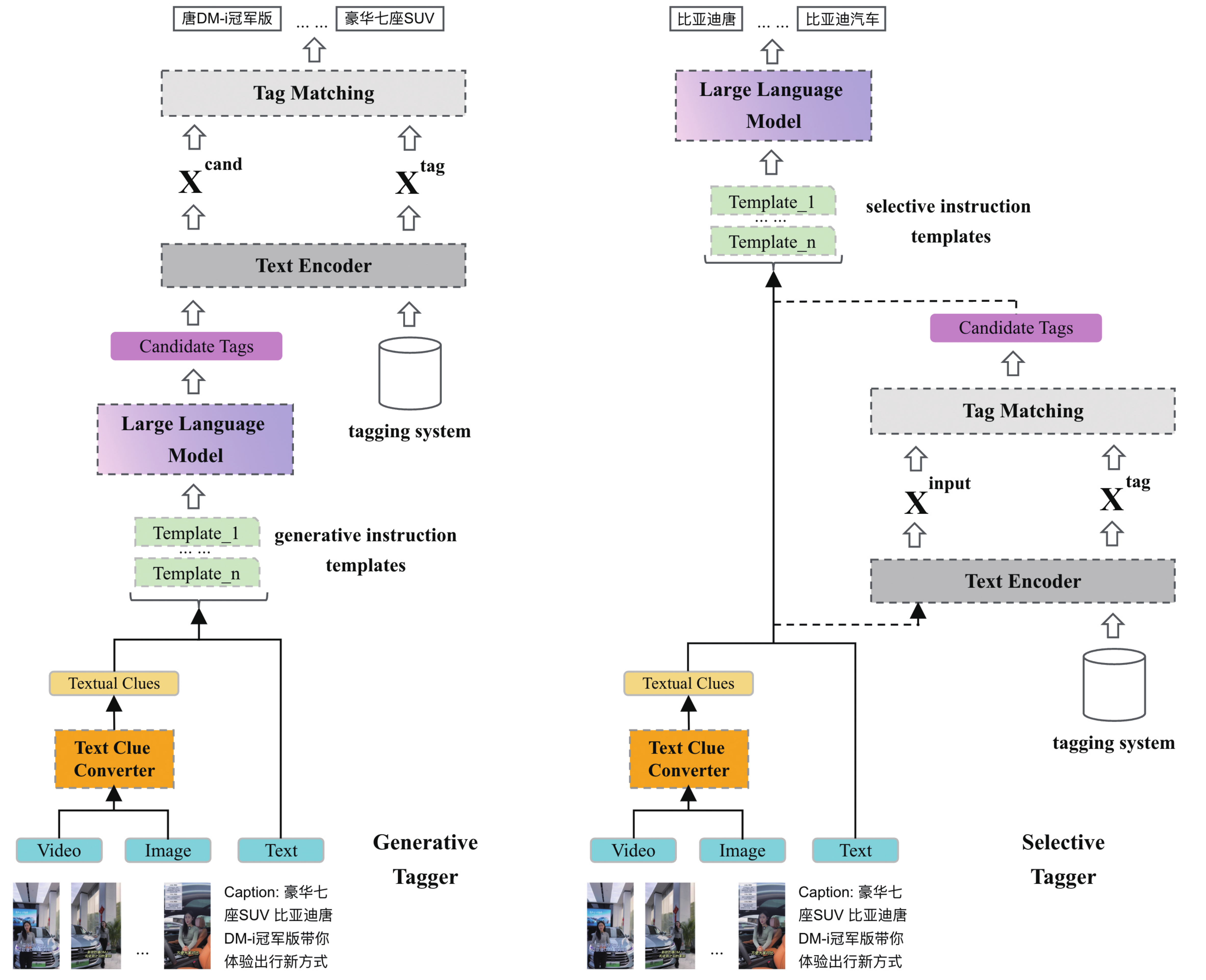}
    \caption{Given the tagging system established in Figure \ref{fig:tag_system}, \modelname\ enables zero-shot tagging of new data in two alternative paradigms.}
    \label{fig:tagging_model}
\end{figure}

In addition to automatically constructing a content-oriented tagging system, \modelname also provides corresponding zero-shot taggers. As shown in Fig.~\ref{fig:tagging_model}, according to the logic of annotation, we will introduce two different taggers separately.

\subsubsection{Generative tagger}
Due to its powerful understanding and reasoning abilities, LLM is an excellent zero-shot tagger. The previous tagging system was constructed based on this capability. Consequently, if the tags inferred by LLM can be quickly matched with the existing tagging system for semantic similarity, then zero-shot tagging can be achieved. Building on this idea, we propose a generative tagger that converts input multimodal samples to textual clues, fills in the corresponding positions of candidate instruction templates, and uses LLM to provide candidate tags. Candidate tags are encoded as $\mathbf{X}^{cand} \in \mathbf{R}^{n \times m}$ using a pre-trained text encoder, while tags in existing tagging systems are encoded as $\mathbf{X}^{tag} \in \mathbf{R}^{N \times m}$, $n, N, m$ denote the number of candidate tags, the number of tags in the tagging system and the representation dimension. The matching score matrix $\mathbf{S} \in \mathbf{R}^{N \times n}$ is then calculated as $\mathbf{S}_{i,j} = \mathbf{X}^{tag}_{i} \cdot \mathbf{X}^{cand \top}_{j} / (|\mathbf{X}^{tag}_{i}| \times |\mathbf{X}^{cand}_{j}|)$. Finally, the matching result can be directly selected through the preset threshold. The left part of Fig.~\ref{fig:tagging_model} illustrates this process.

\subsubsection{Selective tagger}
When LLM is put into practice, people discover its strong ability to understand context, and some methods even use in-context learning strategies to guide LLM in producing more reasonable outcomes~\cite{min2022rethinking}. This inspired us to develop another zero-shot tagging paradigm, as depicted in the right part of Fig.~\ref{fig:tagging_model}. Specifically, we first convert the given multimodal samples into textual clues. Unlike the generative tagger, we input it along with the tags in the tagging system into a pre-trained text encoder to obtain the corresponding vector representation, \textit{i.e.}, $\mathbf{X}^{input}$ and $\mathbf{X}^{tag}$. By matching similarities between the text and tags, we can obtain a series of related candidate tag sets. Finally, by incorporating the original textual clues and candidate tag sets into the selective instruction template and feeding them into LLM, we can obtain the final annotation result for a given multimodal sample.

%% file: texts/3.experiments.tex
\section{Experiments}\label{sec:exp}
In this section, we will test \modelname's ability to construct the tagging system and assign tags to the given multimodal data in the zero-shot setting on two real-world social network datasets.

\subsection{Dataset and Metrics}
\begin{table}[ht]
    \centering
    \begin{tabular}{c|ccccc}
    \toprule
    Dataset & \# items & caption & category & OCR & ASR \\
    \midrule
    Kuaishou & 222k & \ding{52} & \ding{52} & \ding{52} & \ding{52} \\
    Food.com & 5.4k & \ding{52} & \ding{52} & \ding{55} & \ding{52} \\
    \bottomrule
    \end{tabular}
    \vspace{0.1in}
    \caption{The statistics of the datasets.}
    \label{tab:dataset}
\end{table}

We have chosen two renowned websites, Kuaishou\footnote{https://www.kuaishou.com/}~\cite{nie2022search} and Food.com\footnote{https://www.food.com/}, as the data sources for our experiment. The detailed statistics are presented in Tab.~\ref{tab:dataset}.

When evaluating the \modelname-based tagging system, we will use the source's own hashtag system (\textit{i.e.}, extracted from Kuaishou and food.com) as the baseline. To comprehensively evaluate the tagging system's quality, we will refer to previous research designs and consider the following metrics:
\begin{itemize}
    \item \textbf{Popularity}: When a tag is consistently assigned to a group of related objects by the majority of users, it indicates that the tag holds stable value and reduces the likelihood of it being considered spam. However, if the tag is assigned too broadly (\textit{i.e.}, too popular), its ability to distinguish between objects is greatly diminished, which defeats the purpose of the tag. To assess the indicator, we intend to track the distribution of tags for both the baseline and the \modelname-based tagging system using identical data. This will enable us to observe the deviation in tag popularity between the two tagging systems.
    \item \textbf{Practicality}: This metric includes two sub meanings, \textit{i.e.} \textit{least effort} and \textit{high coverage of multiple facets}. This pair of metrics are related, with the first one (\textit{i.e.}, least effort) focusing on the number of tags assigned to a given object, while the second one focuses on the facet coverage of a given object. In fact, these two indicators are theoretically positively correlated, that is, when fewer tags are assigned, the number of facets they can cover decreases. In this paper, we intend to evaluate this pair of indicators by counting the number of tags in each given data under different tagging systems and assessing the semantic redundancy of tags within each given data. Among them, the semantic redundancy of tags comes from the proportion of similar semantics in the corresponding tags for each given data.
    \item \textbf{Uniformity}: Tags may diverge due to personal habits and application scenarios, resulting in different descriptions of the same concept. While keeping these contents can improve the recall rate of annotations and enhance the user experience, excessive tags will increase the tagging system size and introduce noise. When evaluating, we will consider the semantic redundancy (\textit{i.e.}, semantic similarity) between any two tags in the whole tagging system as the calculation method.
\end{itemize}

\subsection{Setup}
In the selection of LLM, we chose GPT 3.5~\cite{instructgpt2022training} (API interface, the version is \textit{gpt-3.5-turbo}). The model of choice for the unsupervised text encoder is SimCSE~\cite{simcse2021emnlp}. In the judgment of semantic redundancy, we take the cosine similarity of the unsupervised semantic vector representation as the measure index. When the similarity is greater than or equal to 0.8, we consider that the two input objects have non-negligible semantic redundancy.

To be able to set a reasonable baseline for comparison for \modelname, we take the tags provided by users when they upload videos for themselves as the baseline. Specifically, we extract the content of the given data spaced by "\#" as the result of the user tagger and collect them as the tagging system of the user annotation. In addition, we also select two well-known visual\footnote{Azure Cognitive Services for Vision (ACSV): https://portal.vision.cognitive.azure.com/demo/generic-image-tagging.} and text\footnote{Hanlp: https://www.hanlp.com/demonstrate.html.} online tagging methods for comparison.

\subsection{Main Result}
In this section, we will evaluate the performance of different sub-modules in \modelname from the two aspects of the tagging system and taggers respectively.

\subsubsection{Tagging system}

\begin{figure}[ht]
    \centering
    \subfigure[Kuaishou] {\includegraphics[width=.49\linewidth]{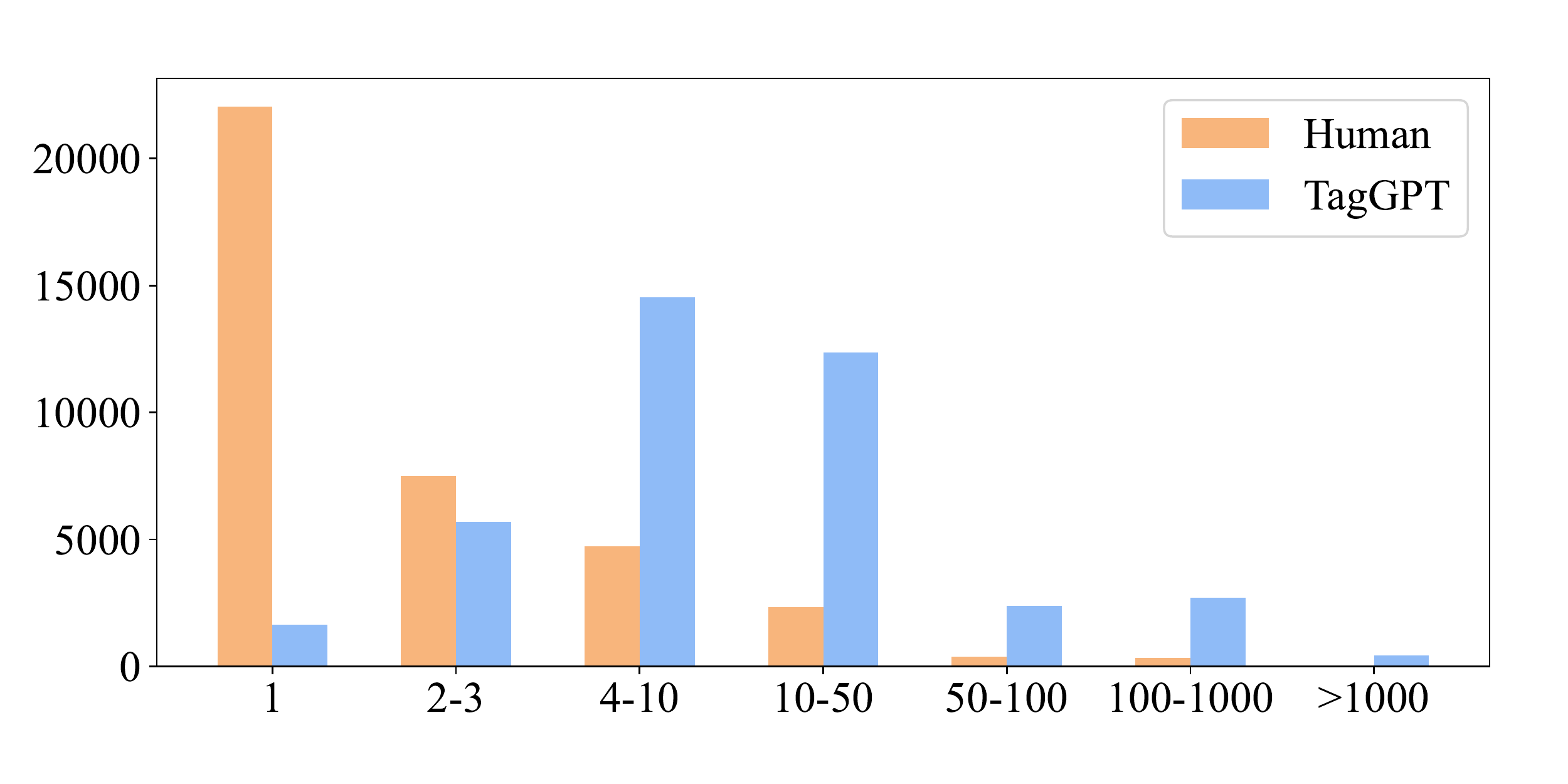}}
    \subfigure[Food.com] {\includegraphics[width=.49\linewidth]{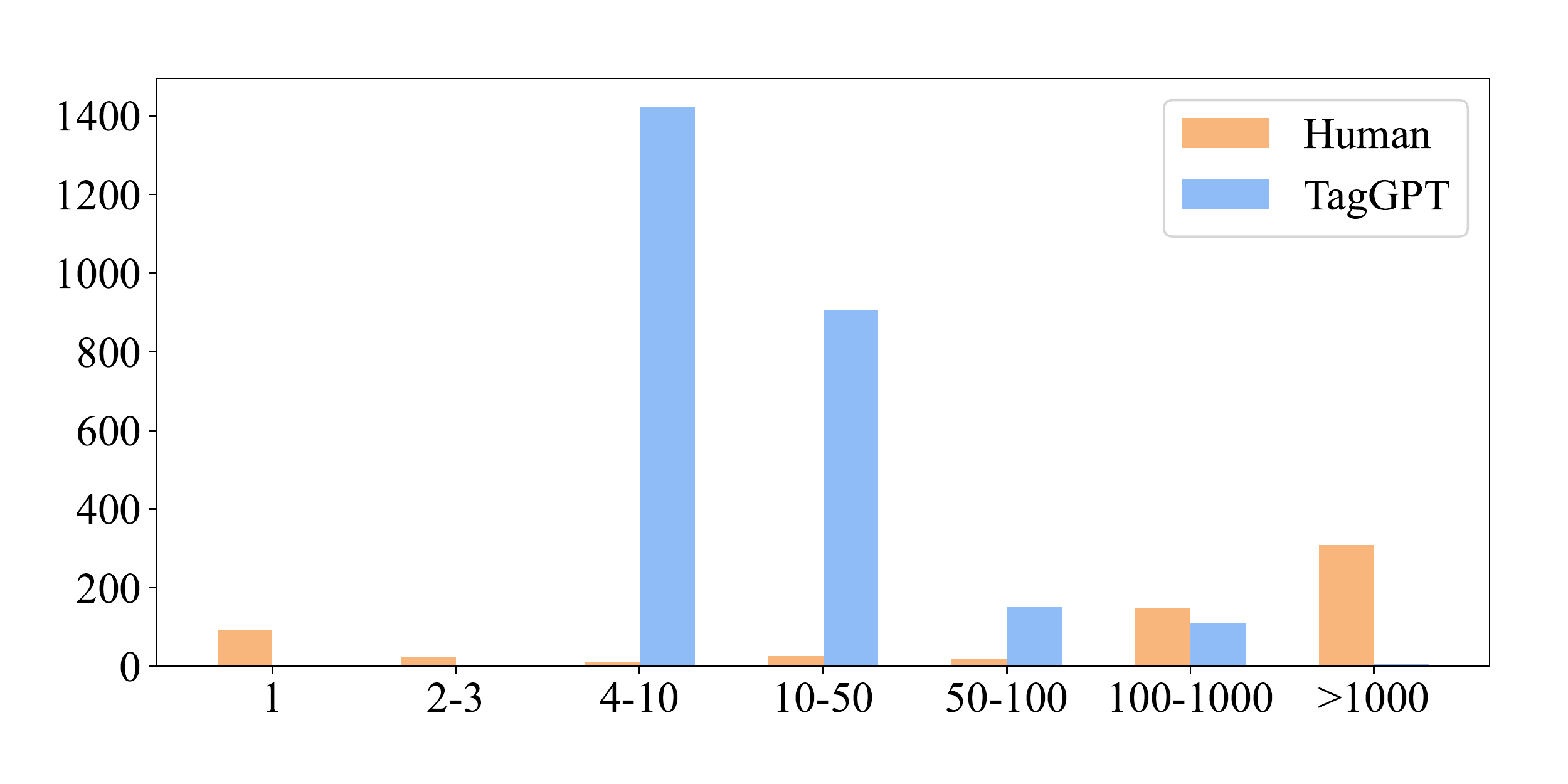}}
    \caption{Statistical results of the metric ``popularity'' in the datasets. The horizontal axis denotes the number of times a tag is assigned to the data, and the vertical axis denotes the number of tags.}
    \label{fig:metric_popularity}
\end{figure}

Fig~\ref{fig:metric_popularity} presents the statistical results of \modelname's tag distribution on the ``popularity'' metric. It is evident that the distribution of tags annotated by users is not optimal in either of the datasets. In the Kuaishou dataset, each tag is used by the user less than four times, and a staggering 60\% of the tags are only used once. On the other hand, in the Food.com dataset, the tag distribution tends to be polarized, where some tags are used too infrequently while others are used too frequently. As mentioned earlier, if the tag popularity is too low, the system will have to face a large number of useless tags, making it difficult for users to efficiently match the correct information. Conversely, if the tag popularity is too high, the system will struggle to effectively distinguish data, leading to a suboptimal user experience. \modelname exhibits excellent tag coverage in both datasets, with a significant number of tags being reused more than 10 times without being overused. This feature facilitates downstream recommendation, search, and other algorithms, ultimately enhancing their performance.

\begin{figure}[ht]
    \centering
    \subfigure[Kuaishou] {\includegraphics[width=.49\linewidth]{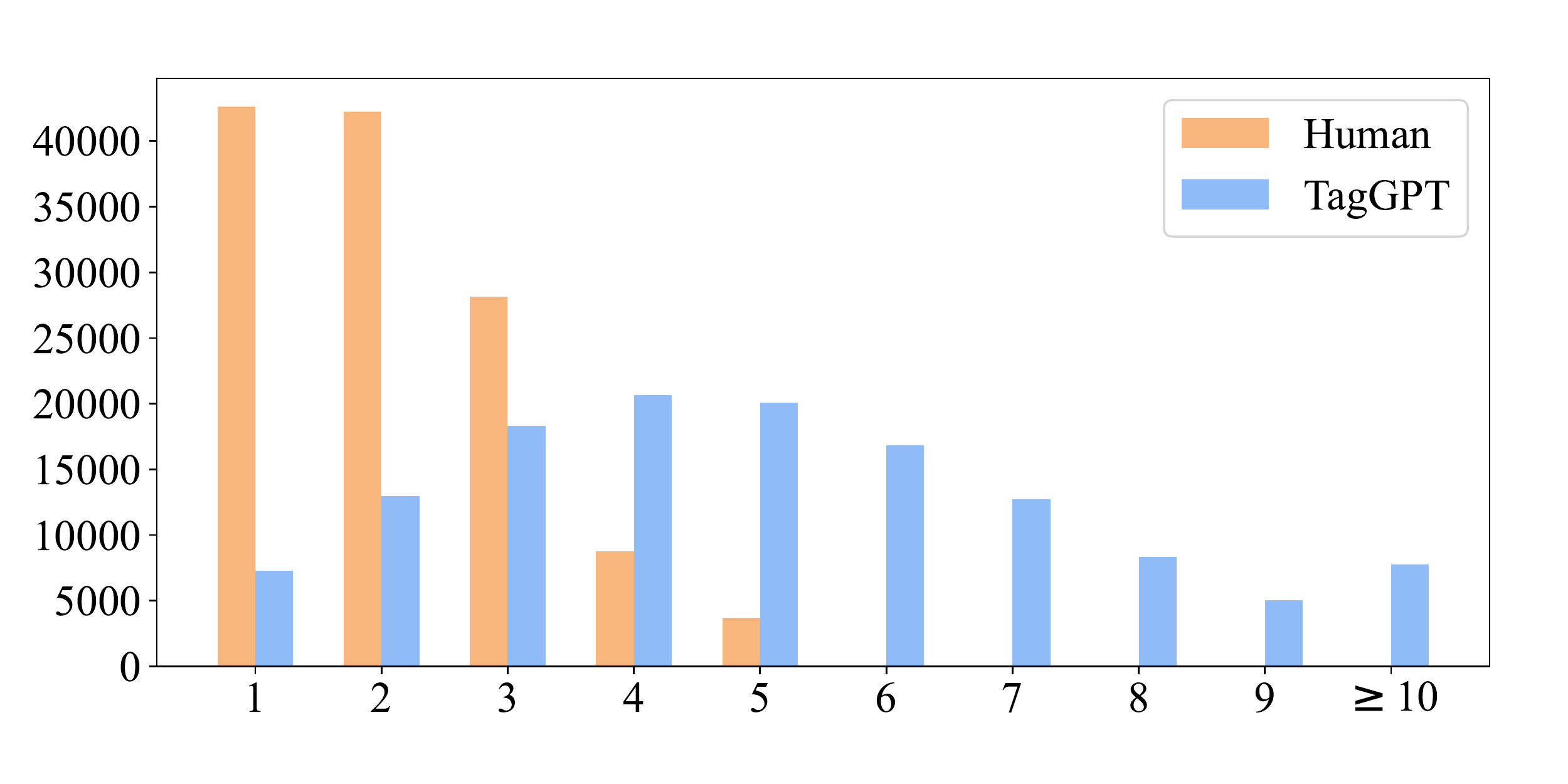}}
    \subfigure[Food.com] {\includegraphics[width=.49\linewidth]{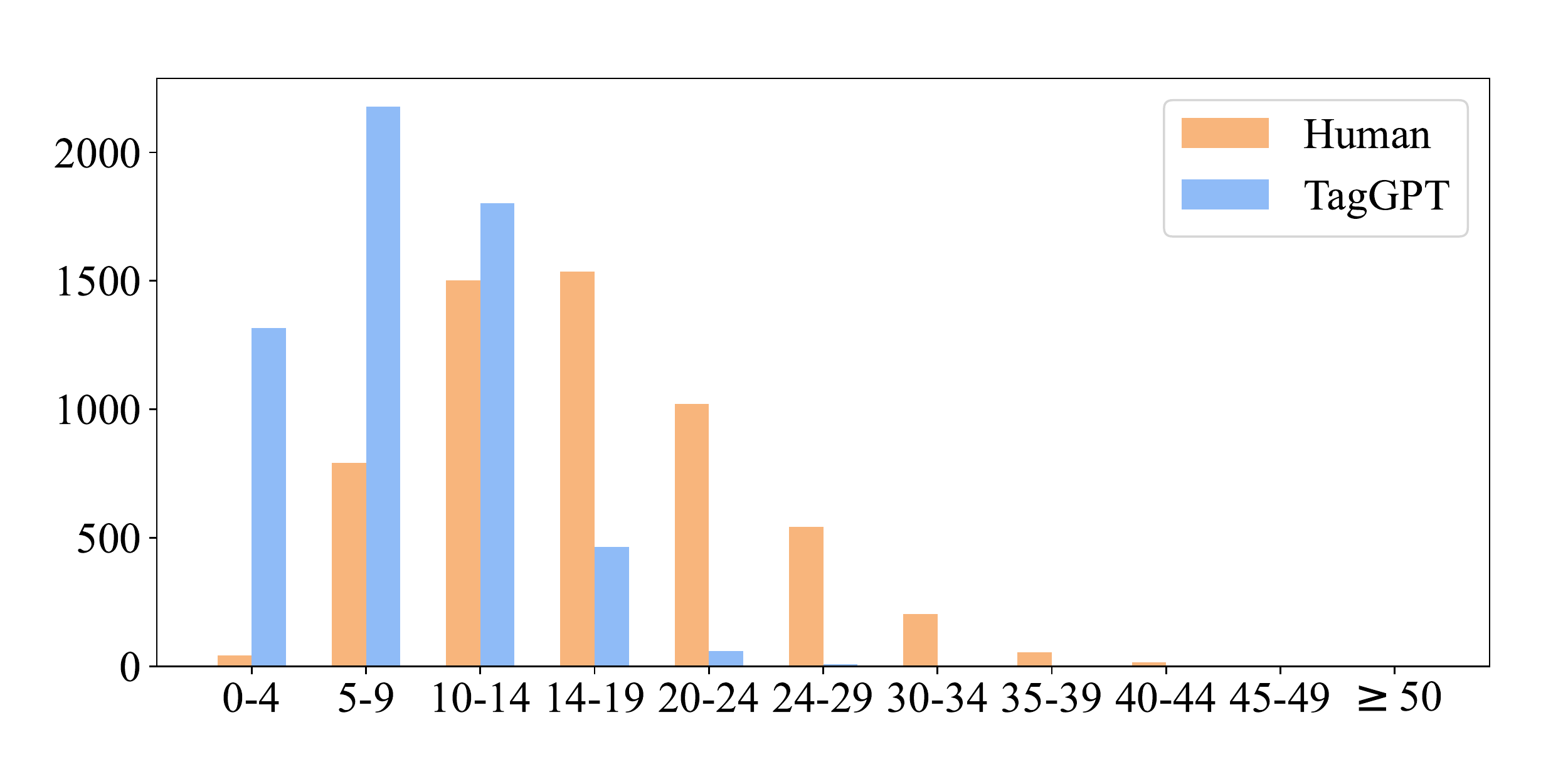}}
    \caption{Statistical results of ``least effort'' in the metric “practicality" in the dataset. The horizontal axis denotes the number of tags assigned to a single data sample, and the vertical axis is the number of samples.}
    \label{fig:metric_least_effort}
\end{figure}
To assess the metric ``practicality'' of tagging systems, we will analyze it from two angles: the number of tags assigned to each sample (i.e., "least effort") and the redundancy of tags within each sample (\textit{i.e.}, ``high coverage of multiple facets''). Fig~\ref{fig:metric_least_effort} displays the tag count for the dataset, revealing that \modelname assigns significantly more tags to each video than the baseline method. Furthermore, we calculated the average tag redundancy rate within each sample for both tagging systems. \modelname's redundancy rate is 0\% in both datasets because it filters out redundant tags during tag statistics. In comparison, the baseline method sets a similarity threshold of 0.8 for statistics, resulting in an average tag redundancy of 5.86\% and 3.77\%. This demonstrates that \modelname not only assigns more tags to each sample than the baseline method but also ensures that the semantic coherence between these tags is low, aligning with the diverse perspectives of each sample.

\begin{table}[t]
    \centering
    \begin{tabular}{c|c|c|c}
        \toprule
        Dataset & Source & \# tags & Redundancy Ratio \\
        \midrule
        \multirow{2}{*}{Kuaishou} & Human & 37,320 & 2.85\% \\ 
         & \modelname & 40,238 & 0.0\% \\ 
        \midrule
        \multirow{2}{*}{Food.com} & Human & 458 & 0.97\% \\
         & \modelname & 2,598 & 0.0\% \\
        \bottomrule
    \end{tabular}
    \vspace{0.1in}
    \caption{``Uniformity'' scores in the corresponding tagging systems for both datasets.}
    \label{tab:metric_uniformity}
\end{table}

When calculating the ``uniformity'' score, we will pair all tags in the tagging system to determine their semantic similarity. We then calculate the proportion of tag pairs whose similarity exceeds the threshold of 0.8 among all tag pairs. The results are presented in Tab.~\ref{tab:metric_uniformity}, showing that the internal uniformity of the tagging system constructed by \modelname is significantly better than the tagging system based on human annotation. This proves that \modelname has lower tag redundancy. Furthermore, when considering the number of tags and their redundancy, \modelname's tags can cover a wider range of content perspectives.

\subsubsection{Taggers}
\begin{table}[t]
    \centering
    \begin{tabular}{c|ccc|ccc}
        \toprule
        \multirow{2}{*}{Dataset} & \multicolumn{3}{c}{selective tagger} & \multicolumn{3}{c}{generative tagger} \\
         & Precision & Recall & F1 & Precision & Recall & F1 \\
        \midrule
        Kuaishou & 78.3 & 69.4 & 73.6 & 80.1 & 72.1 & 75.9 \\
        Food.com & 81.5 & 75.0 & 78.1 & 83.7 & 76.9 & 78.7 \\
        \bottomrule
    \end{tabular}
    \vspace{0.1in}
    \caption{Quantitative results of two alternative taggers in \modelname.}
    \label{tab:tagger}
\end{table}

\begin{figure}[t]
    \centering
    \subfigure[High-frequency tags inferred from the Kuaishou dataset.] {\includegraphics[width=.99\linewidth]{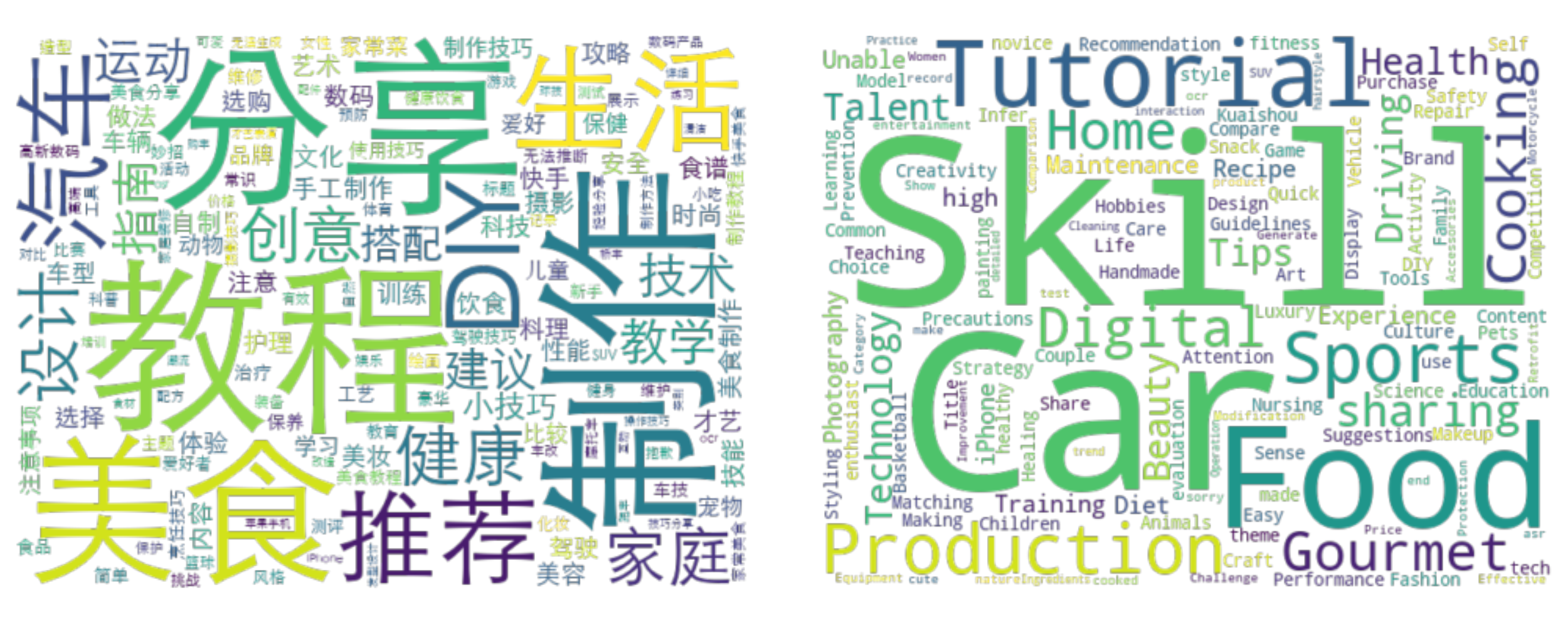}}
    \subfigure[High-frequency tags inferred from the Food.com dataset.] {\includegraphics[width=.99\linewidth]{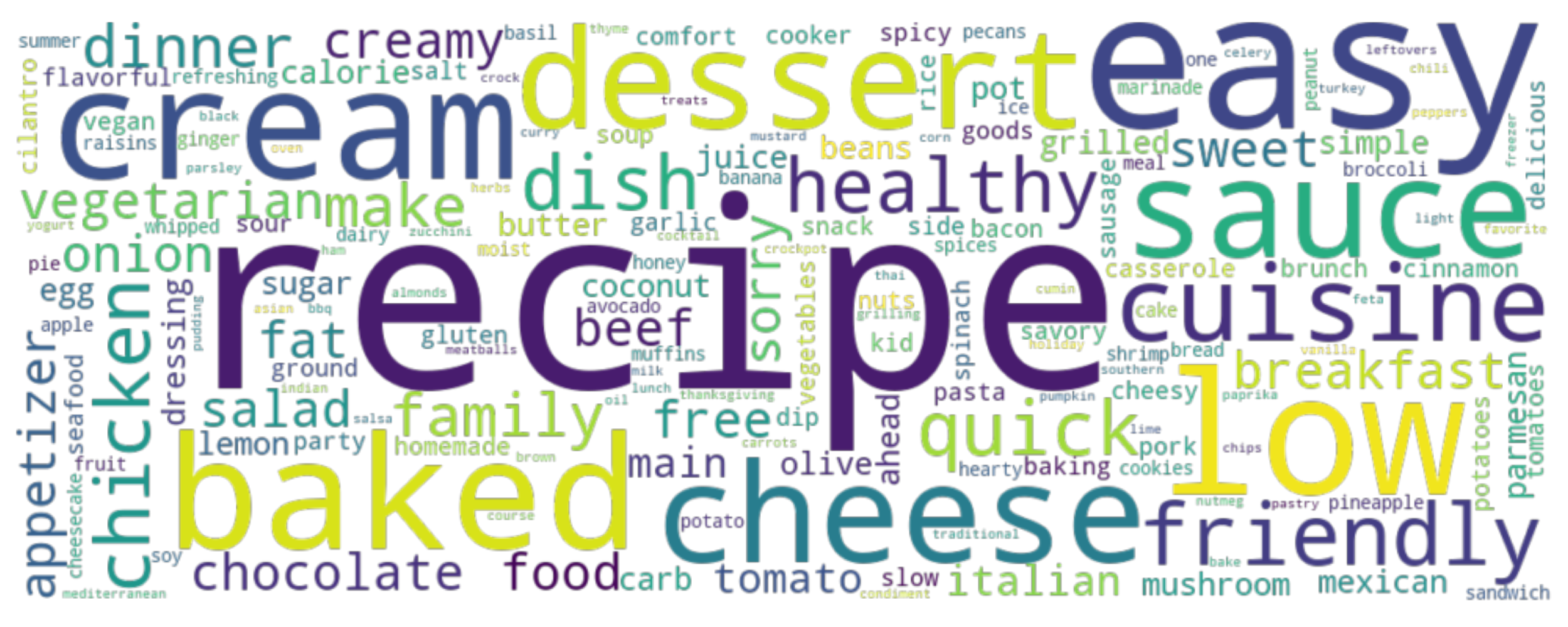}}
    \caption{The word cloud of the top 150 words in the tagging system constructed by \modelname in both datasets.}
    \label{fig:wordcloud_chn}
\end{figure}

In this paper, we introduce two distinct zero-shot taggers, namely generative tagger and selective tagger. To assess the effectiveness of the taggers, we randomly set aside 100 samples from the dataset as a test set, which were not involved in the development of the tagging system. The performance of the taggers was evaluated through manual assessment, and the results are presented in Tab.~\ref{tab:tagger}. Both taggers are proficient in assigning high-quality tags for the given data in a zero-shot setting. A closer comparison of the results of the two paradigms shows that the selective tagger performs slightly worse than the generative tagger, mainly due to the different stages of tag matching. The selective tagger takes all textual clues as input to perform coarse tag matching before the LLM, which may lead to too many input features and exclude some correct fine-grained tags from the candidate set. However, the generative tagger can effectively retain some detail tags, so it is bound to be more in line with the ground truth.

\subsection{Case Study}

To provide a more intuitive understanding of \modelname's performance in zero-shot tagging, this section will present additional case studies to illustrate qualitative results. The specific outcomes are displayed in Tab.~\ref{tab:case_studies}.

Upon comparing the tagging results from various sources in Tab.~\ref{tab:case_studies}, it becomes apparent that human tagger often assigns relatively vague and limited tag semantics when dealing with multimodal data. This is usually because micro video creators cannot quickly and accurately find the semantic tags they need in the massive semantic tags when uploading their works. Moreover, since the popularity of tags directly affects the recommendation algorithm in micro video platforms, creators usually tend to choose popular but ambiguous tags to boost the exposure of their works and reach a broader audience. 

\begin{table}[H]
    \centering
    \tiny
    \begin{tabular}{ccccccc}
    \toprule
    \multicolumn{2}{c}{Key Frames} & Human Tagger & Selective Tagger & Generative Tagger & ACSV & Hanlp \\
    \midrule
    \begin{minipage}{0.06\textwidth}\includegraphics[width=1.1\linewidth]{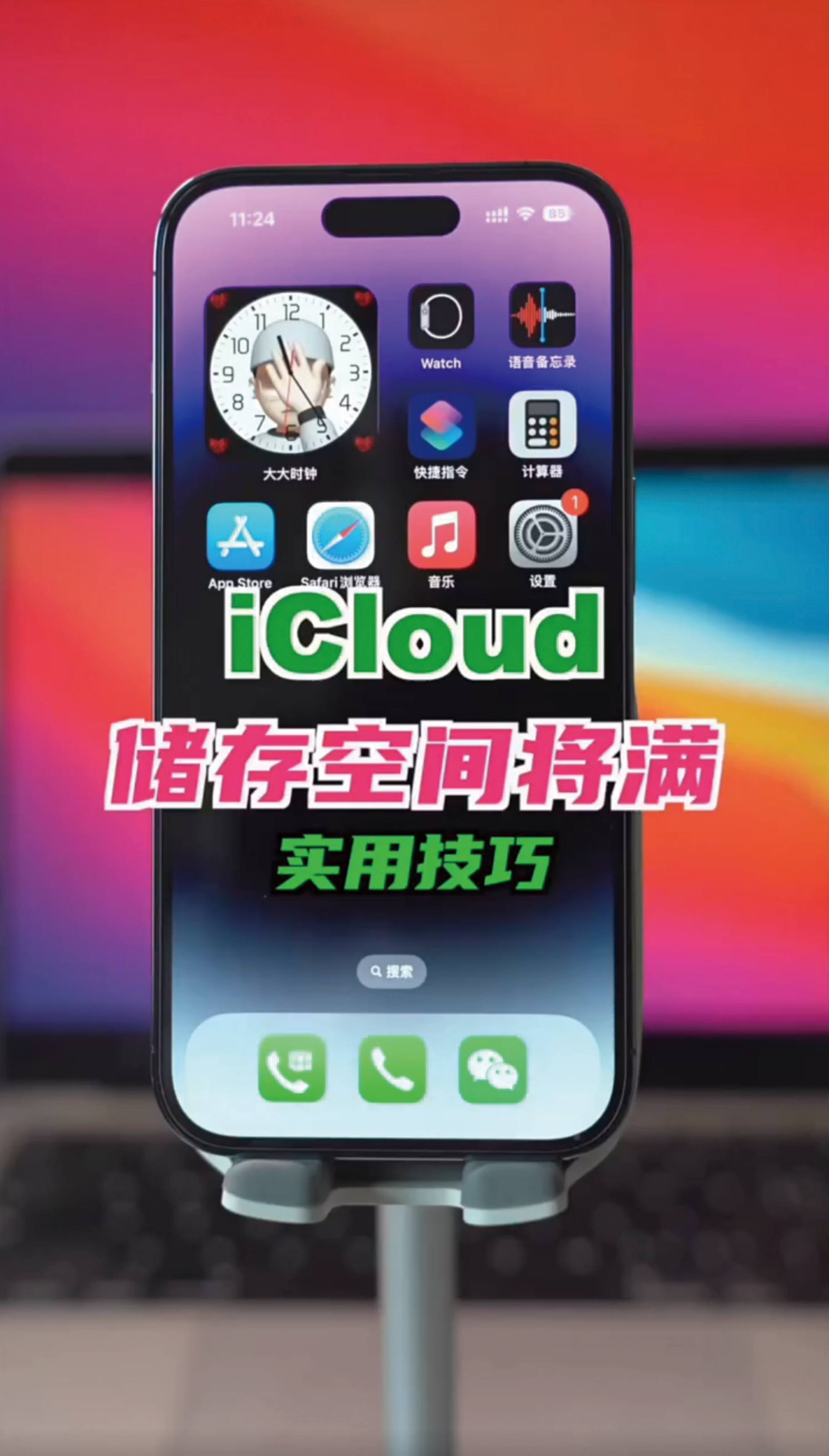}\end{minipage} & \begin{minipage}{0.06\textwidth}\includegraphics[width=1.1\linewidth]{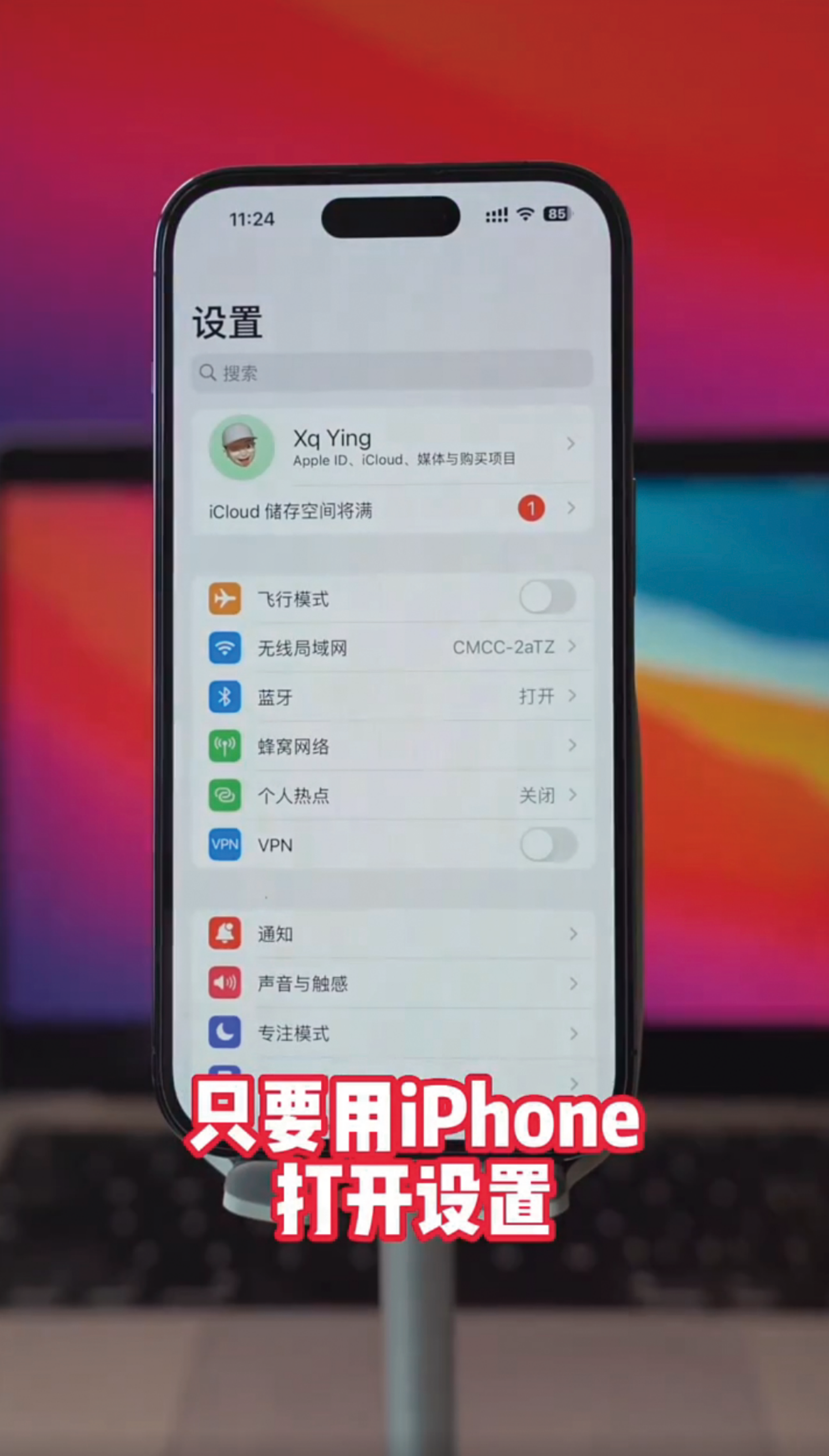}\end{minipage} & \begin{CJK*}{UTF8}{gbsn}\makecell{iPhone小技巧,\\iCloud空间不足,\\苹果手机}\end{CJK*} & \begin{CJK*}{UTF8}{gbsn}\makecell{存储空间管理,\\苹果设备管理,\\iCloud空间不足,\\高新数码产品,\\iPhone常见问题解决}\end{CJK*} & \begin{CJK*}{UTF8}{gbsn}\makecell{iCloud空间管理,\\苹果手机技巧,\\苹果设备管理,\\iCloud存储空间,\\iCloud备份与恢复}\end{CJK*} & \begin{CJK*}{UTF8}{gbsn}\makecell{文本,\\手机,\\屏幕截图,\\小工具,\\多媒体}\end{CJK*} & \begin{CJK*}{UTF8}{gbsn}\makecell{云备份,\\储存空间,\\实用技巧}\end{CJK*} \\
    & & & & & \\
    \begin{minipage}{0.06\textwidth}\includegraphics[width=1.1\linewidth]{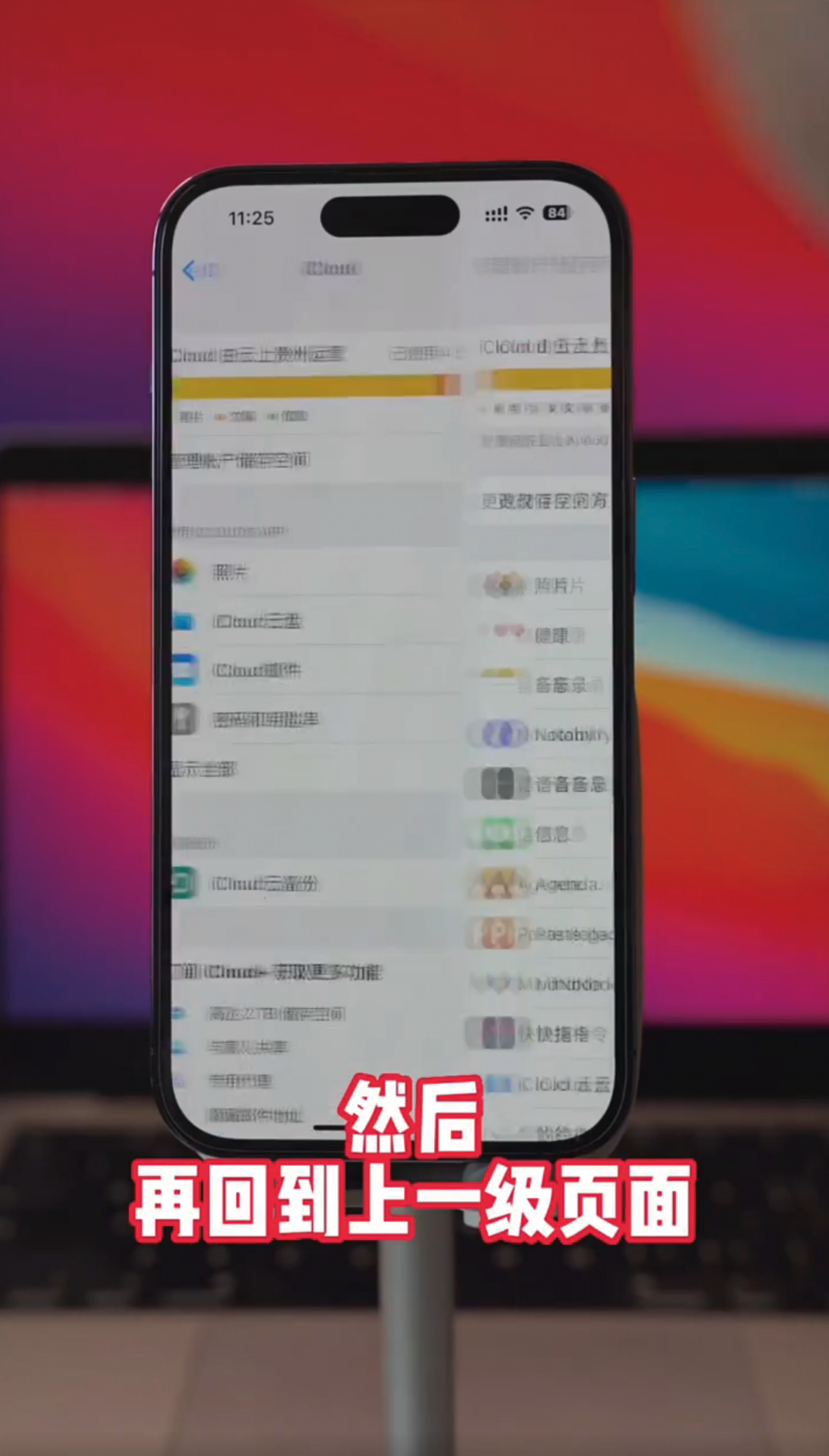}\end{minipage} & \begin{minipage}{0.06\textwidth}\includegraphics[width=1.1\linewidth]{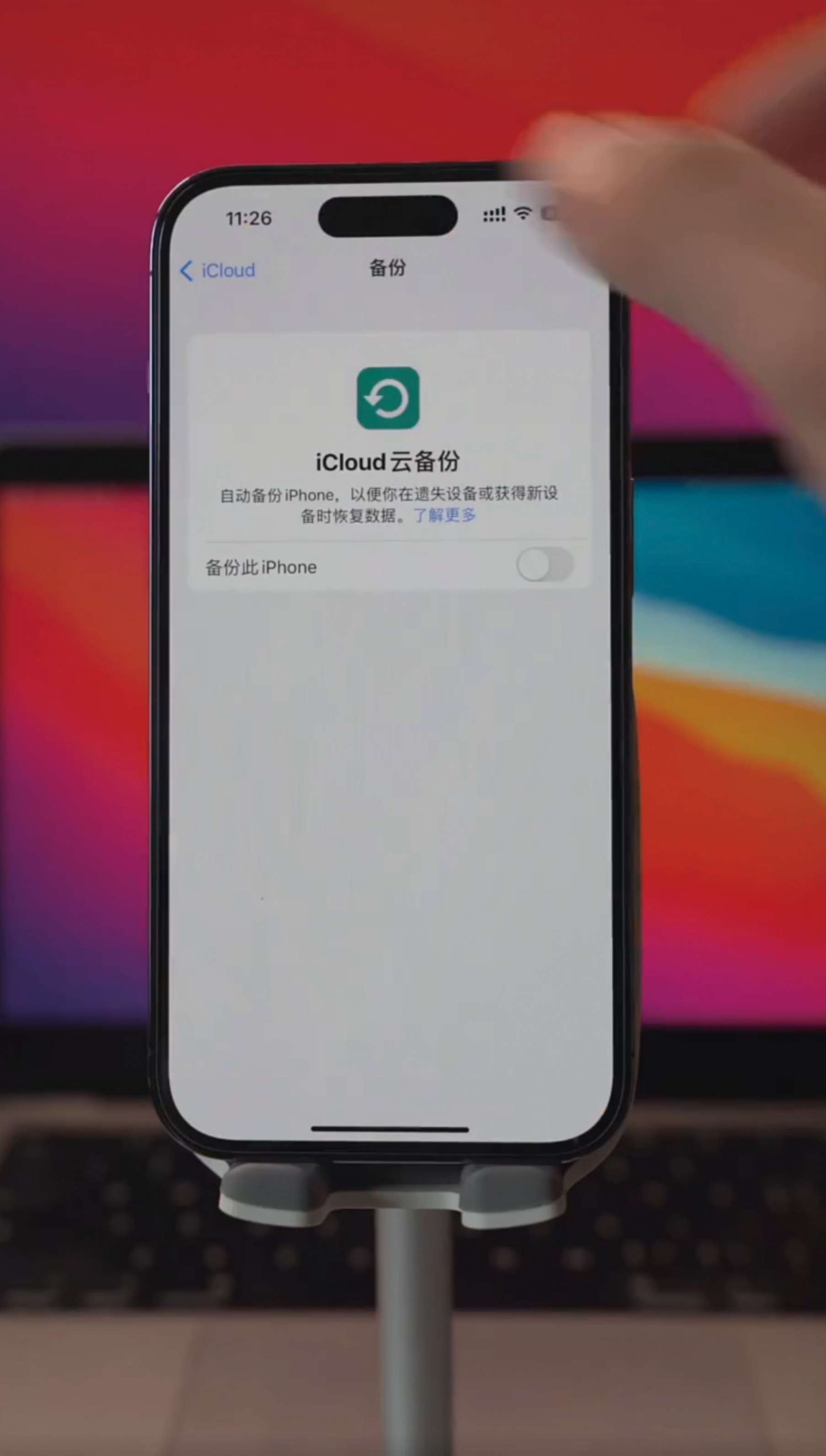}\end{minipage} & \makecell{iPhone tips, \\insufficient\\iCloud\\space, iPhone} & \makecell{storage space management,\\Apple device management,\\iCloud insufficient space,\\high-tech digital products,\\iPhone common problem solving} & \makecell{iCloud space management,\\Apple mobile phone skills,\\Apple device management,\\iCloud storage space,\\iCloud backup and recovery} & \makecell{text,\\mobile,\\screenshot,\\gadgets,\\multimedia} & \makecell{cloud backup,\\storage space,\\practical skills}\\
    \midrule
    \begin{minipage}{0.06\textwidth}\includegraphics[width=1.1\linewidth]{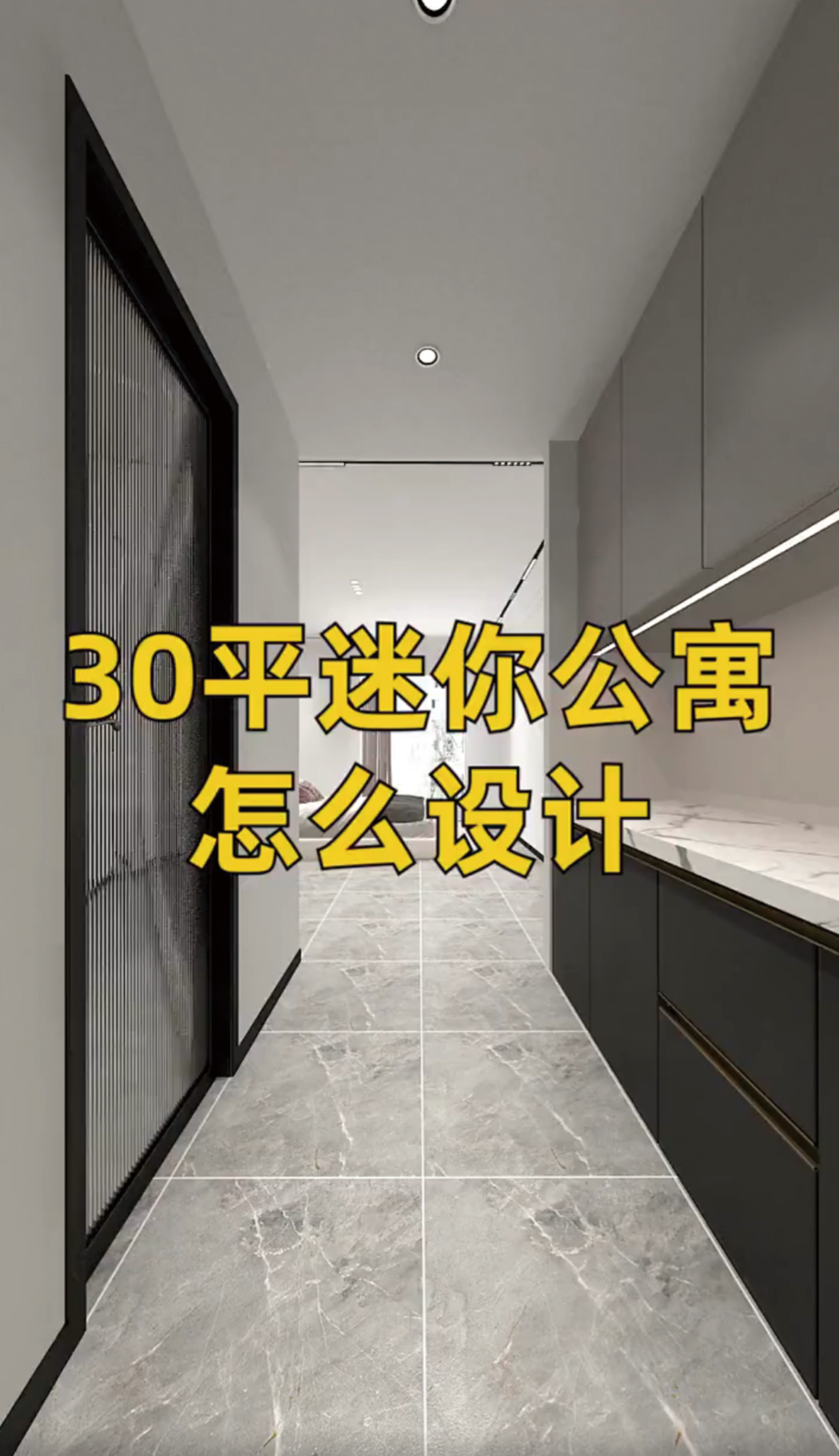}\end{minipage} & \begin{minipage}{0.06\textwidth}\includegraphics[width=1.1\linewidth]{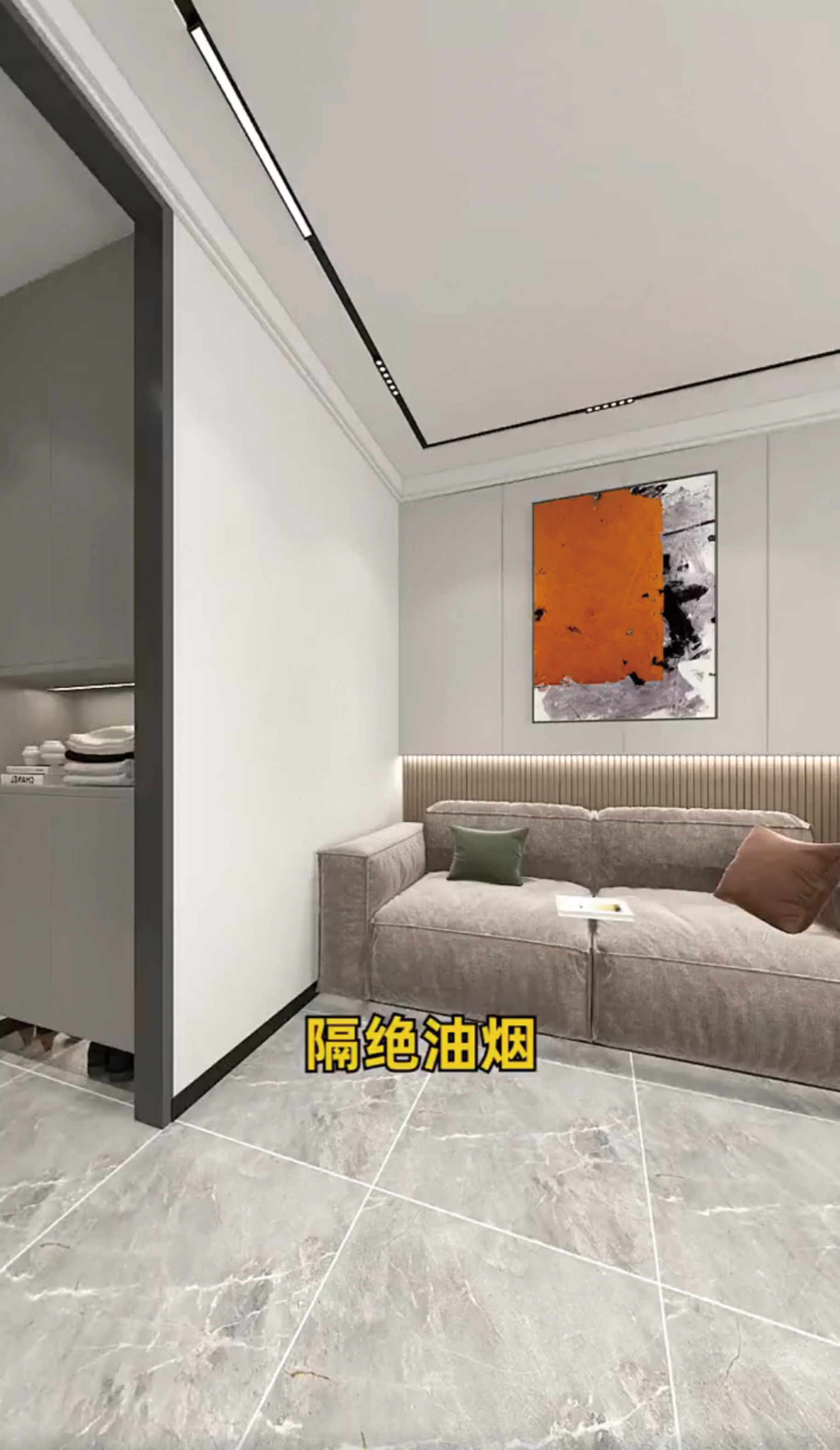}\end{minipage} & \begin{CJK*}{UTF8}{gbsn}\makecell{装修,\\设计,\\公寓装修}\end{CJK*} & \begin{CJK*}{UTF8}{gbsn}\makecell{家居DIY,\\多功能房间设计,\\单身公寓设计,\\小户型设计,\\实用家居}\end{CJK*} & \begin{CJK*}{UTF8}{gbsn}\makecell{迷你公寓设计,\\小户型设计,\\装修技巧,\\创意装修,\\居家便捷设计}\end{CJK*} & \begin{CJK*}{UTF8}{gbsn}\makecell{墙壁,\\室内,\\门,\\瓷砖,\\室内设计}\end{CJK*} & \begin{CJK*}{UTF8}{gbsn}\makecell{鞋柜,\\书柜,\\创意设计,\\公寓}\end{CJK*} \\
    & & & & & \\
    \begin{minipage}{0.06\textwidth}\includegraphics[width=1.1\linewidth]{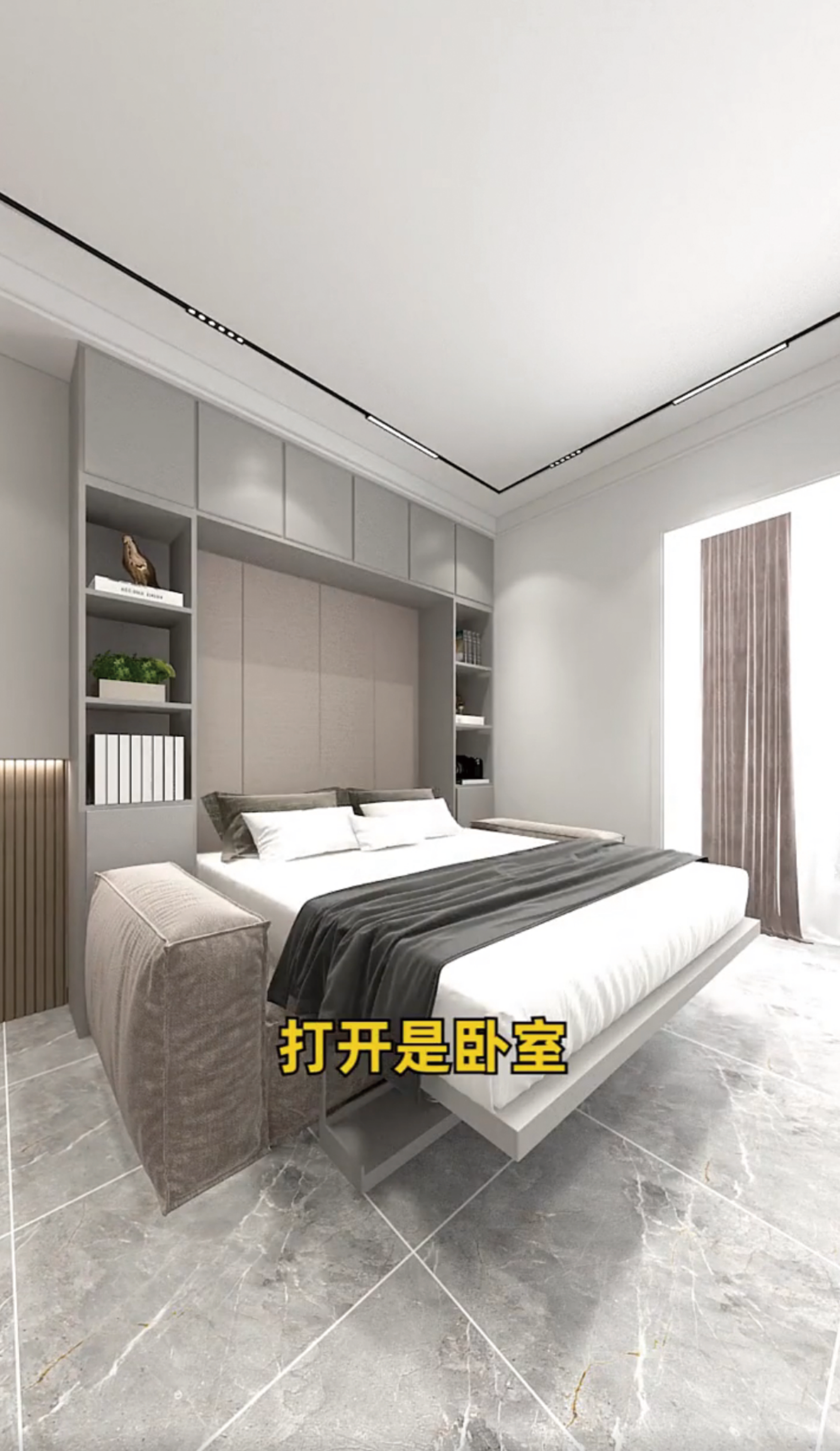}\end{minipage} & \begin{minipage}{0.06\textwidth}\includegraphics[width=1.1\linewidth]{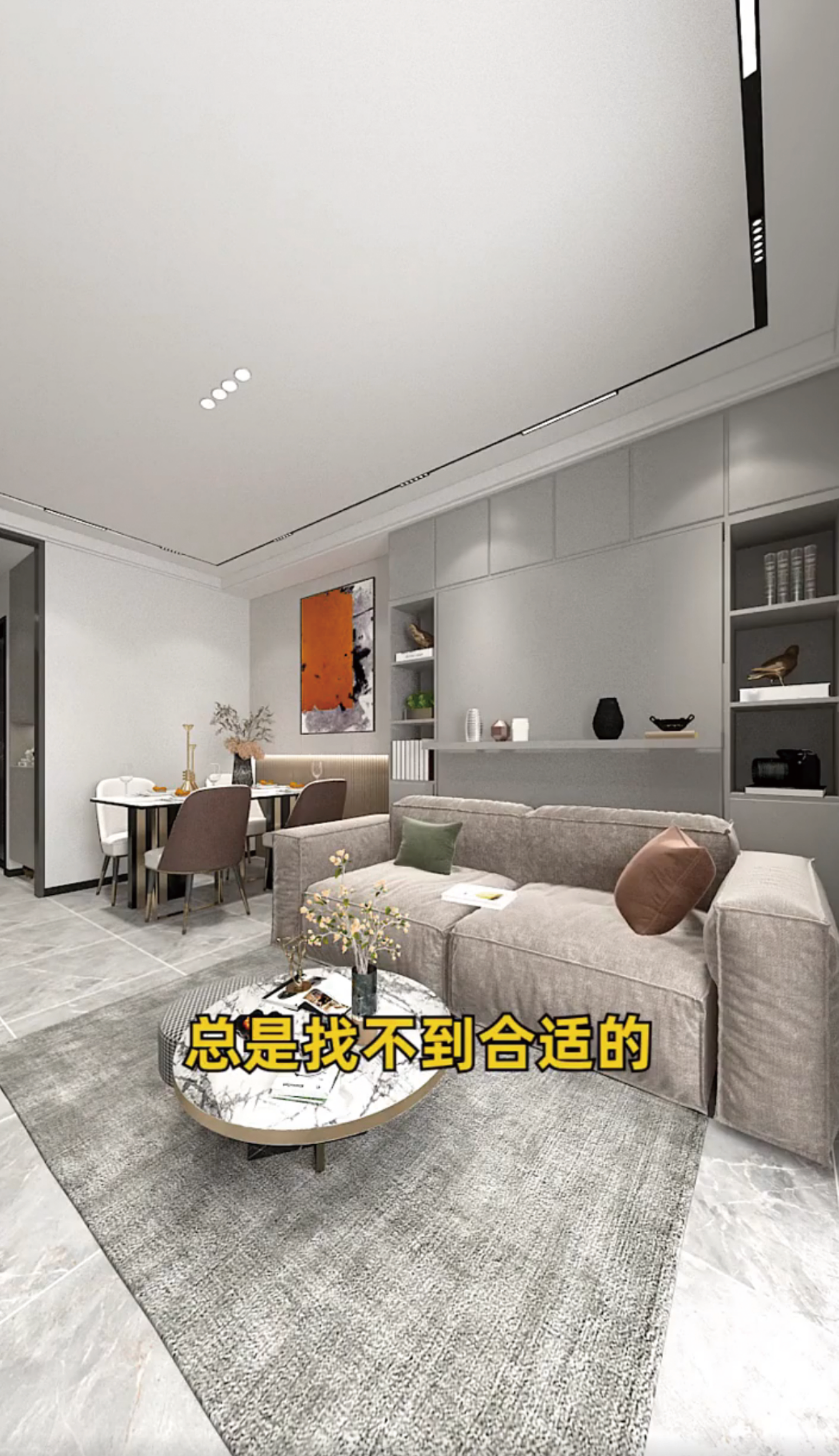}\end{minipage} & \makecell{decoration,\\design,\\apartment\\decoration} & \makecell{home DIY,\\multifunctional room design,\\single apartment design,\\small apartment design,\\practical home furnishing} & \makecell{mini apartment design,\\small apartment design,\\decoration skills,\\creative decoration,\\convenient home design} & \makecell{wall,\\interior,\\door,\\tile,\\interior design}  & \makecell{shoe cabinet,\\bookcase,\\creative design,\\apartment}\\
    \midrule
    \begin{minipage}{0.06\textwidth}\includegraphics[width=1.1\linewidth]{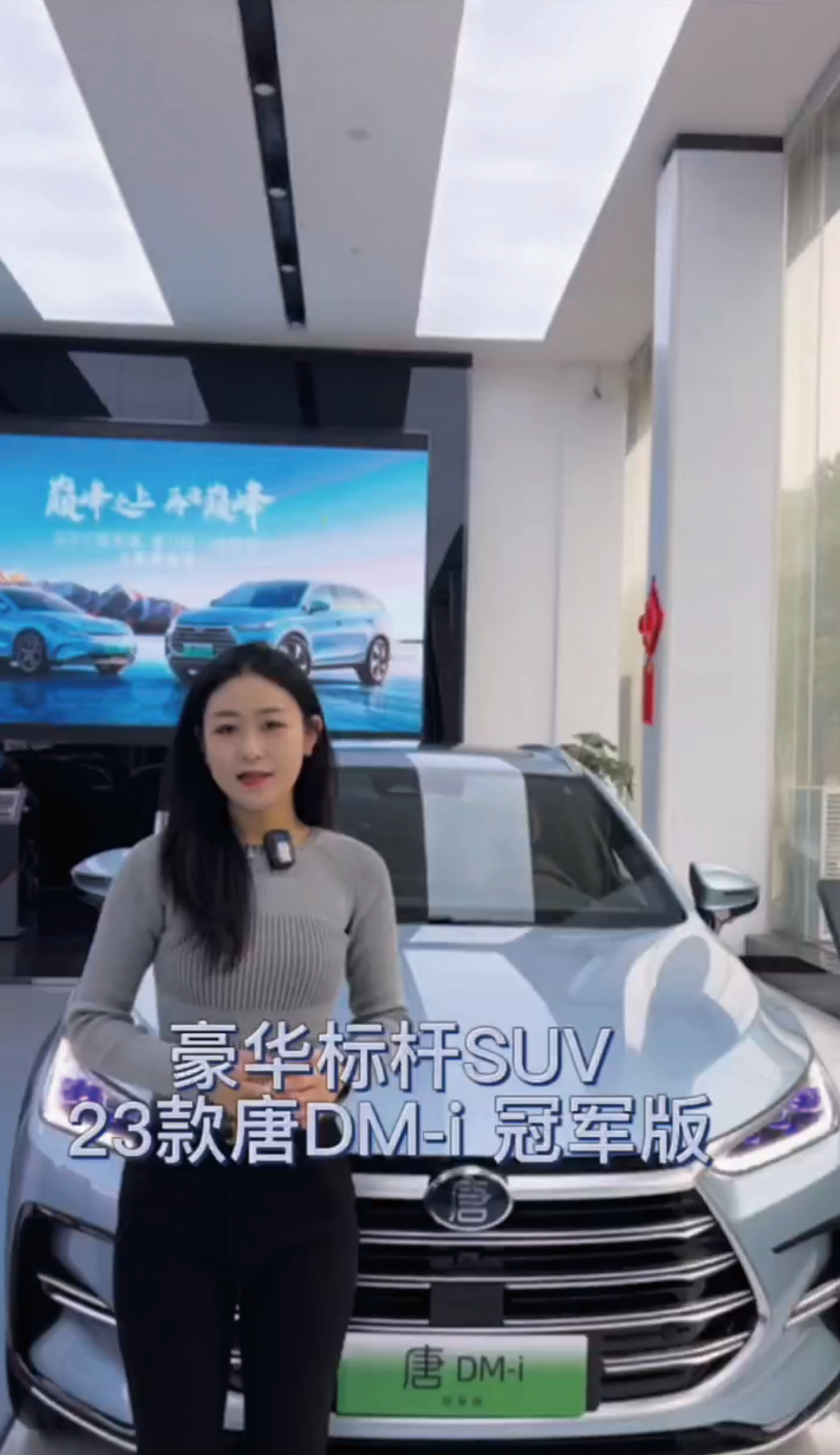}\end{minipage} & \begin{minipage}{0.06\textwidth}\includegraphics[width=1.1\linewidth]{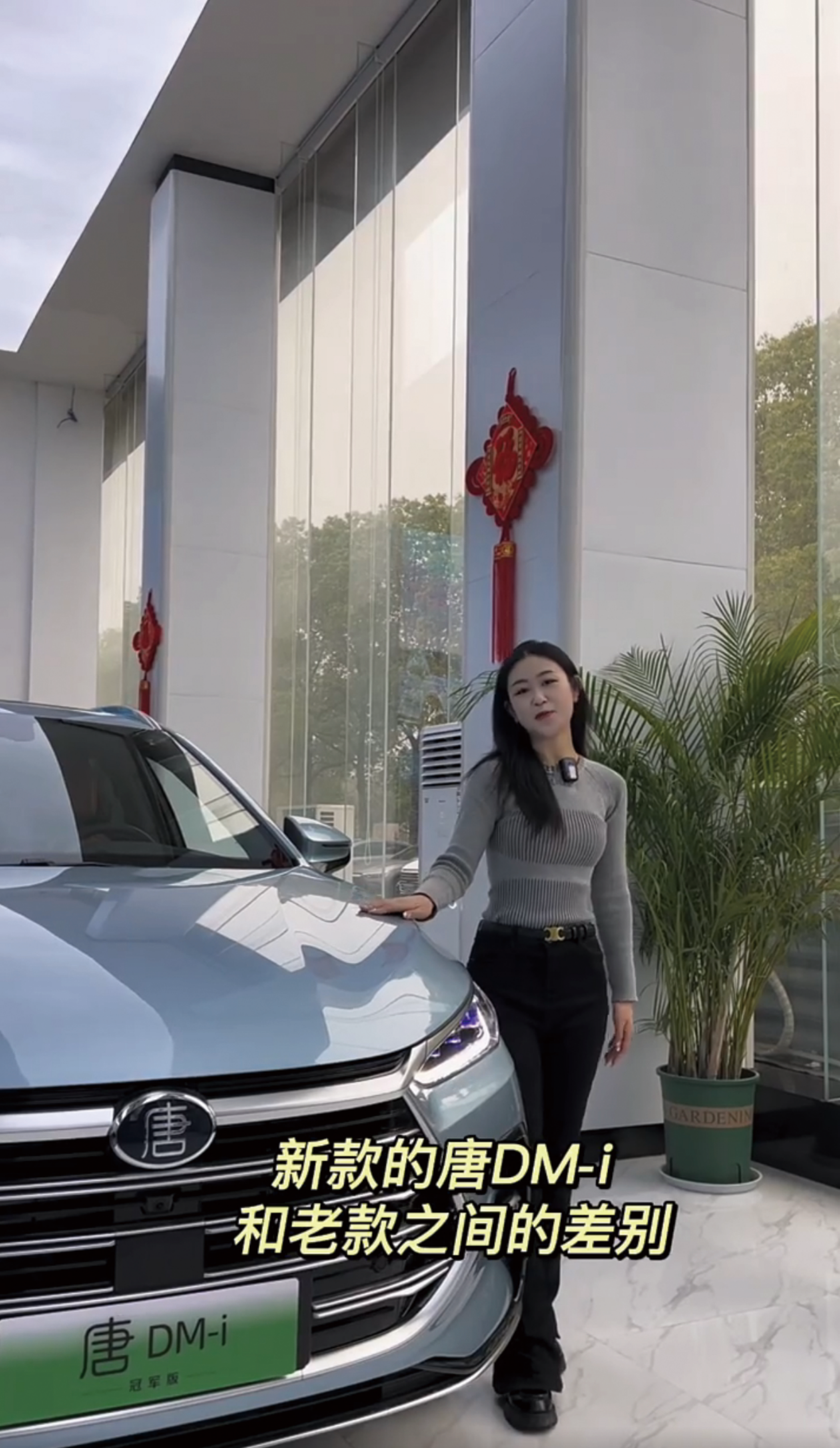}\end{minipage} & \begin{CJK*}{UTF8}{gbsn}\makecell{比亚迪,\\新能源汽车,\\唐dmi冠军版}\end{CJK*} & \begin{CJK*}{UTF8}{gbsn}\makecell{比亚迪唐,\\SUV汽车,\\汽车销量分析,\\豪华七座SUV,\\比亚迪汽车}\end{CJK*} & \begin{CJK*}{UTF8}{gbsn}\makecell{唐DM-i冠军版,\\新能源汽车,\\比亚迪唐DM-i,\\豪华SUV,\\豪华七座SUV}\end{CJK*} & \begin{CJK*}{UTF8}{gbsn}\makecell{车辆,\\陆地车辆,\\文本,\\衣服,\\汽车设计}\end{CJK*} & \begin{CJK*}{UTF8}{gbsn}\makecell{老款,\\冠军版,\\快充,\\尊贵,\\比亚迪}\end{CJK*} \\
    & & & & & \\
    \begin{minipage}{0.06\textwidth}\includegraphics[width=1.1\linewidth]{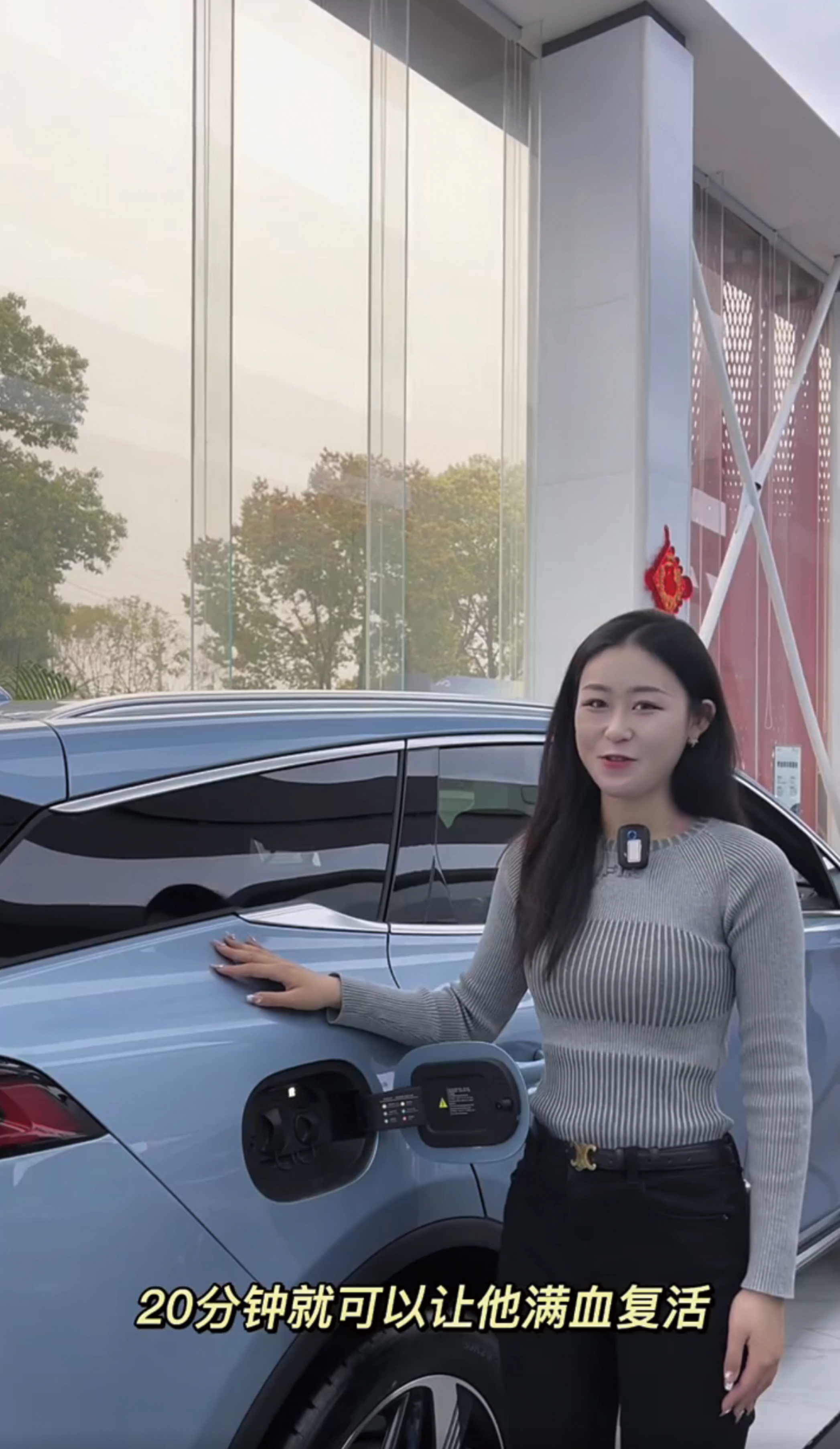}\end{minipage} & \begin{minipage}{0.06\textwidth}\includegraphics[width=1.1\linewidth]{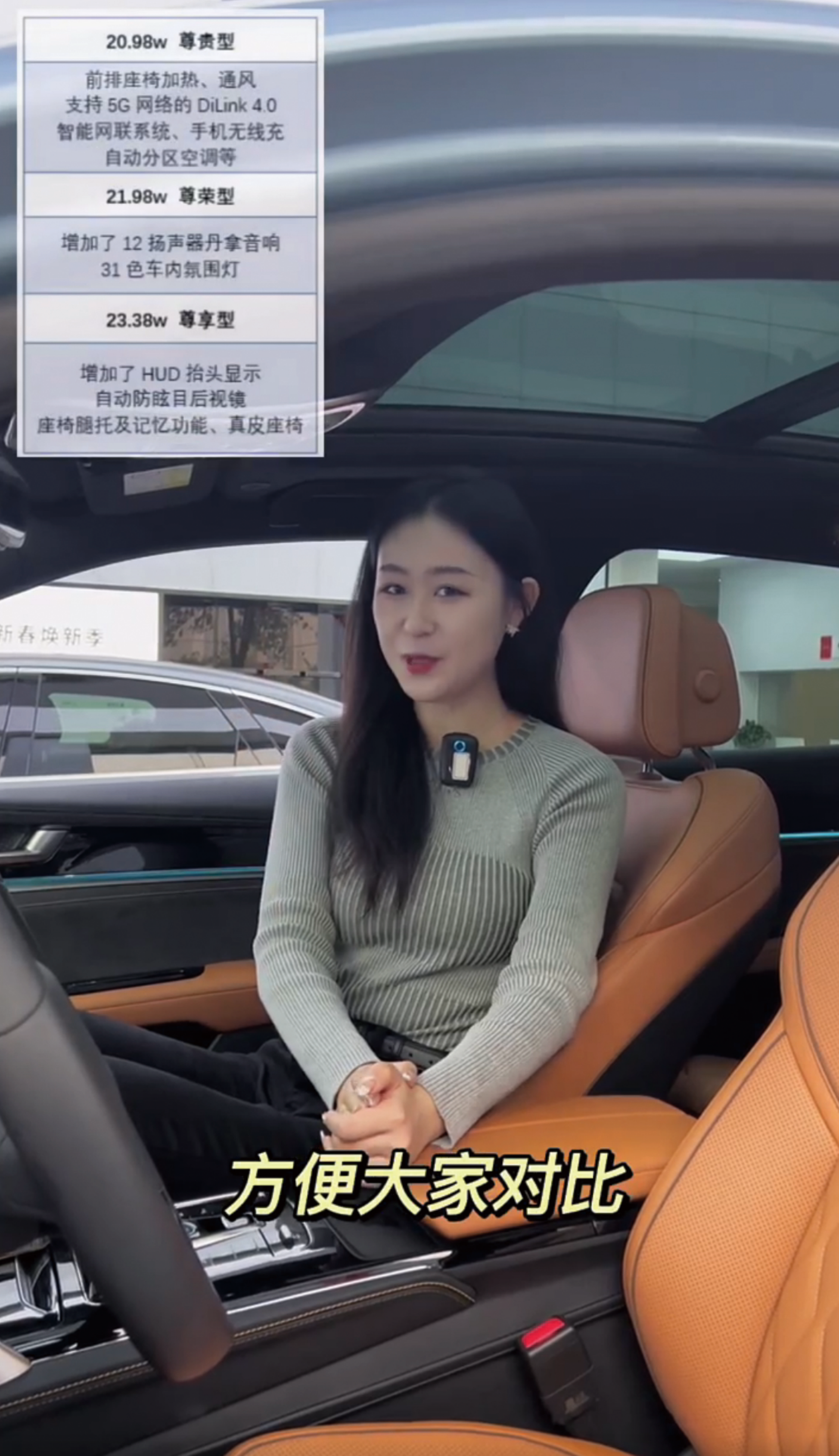}\end{minipage} & \makecell{BYD\\new energy\\vehicle,Tang\\dmi champion\\edition} & \makecell{BYD Tang,\\SUV car,\\car sales analysis,\\luxury seven-seat SUV,\\BYD car} & \makecell{Tang DM-i champion version,\\new energy vehicles,\\BYD Tang DM-i,\\luxury SUV,\\luxury seven-seat SUV} & \makecell{vehicle,\\land vehicle,\\text,\\clothes,\\car design} & \makecell{old model,\\champion version,\\fast charge,\\distinguished,\\BYD} \\
    \midrule
    \begin{minipage}{0.06\textwidth}\includegraphics[width=1.1\linewidth]{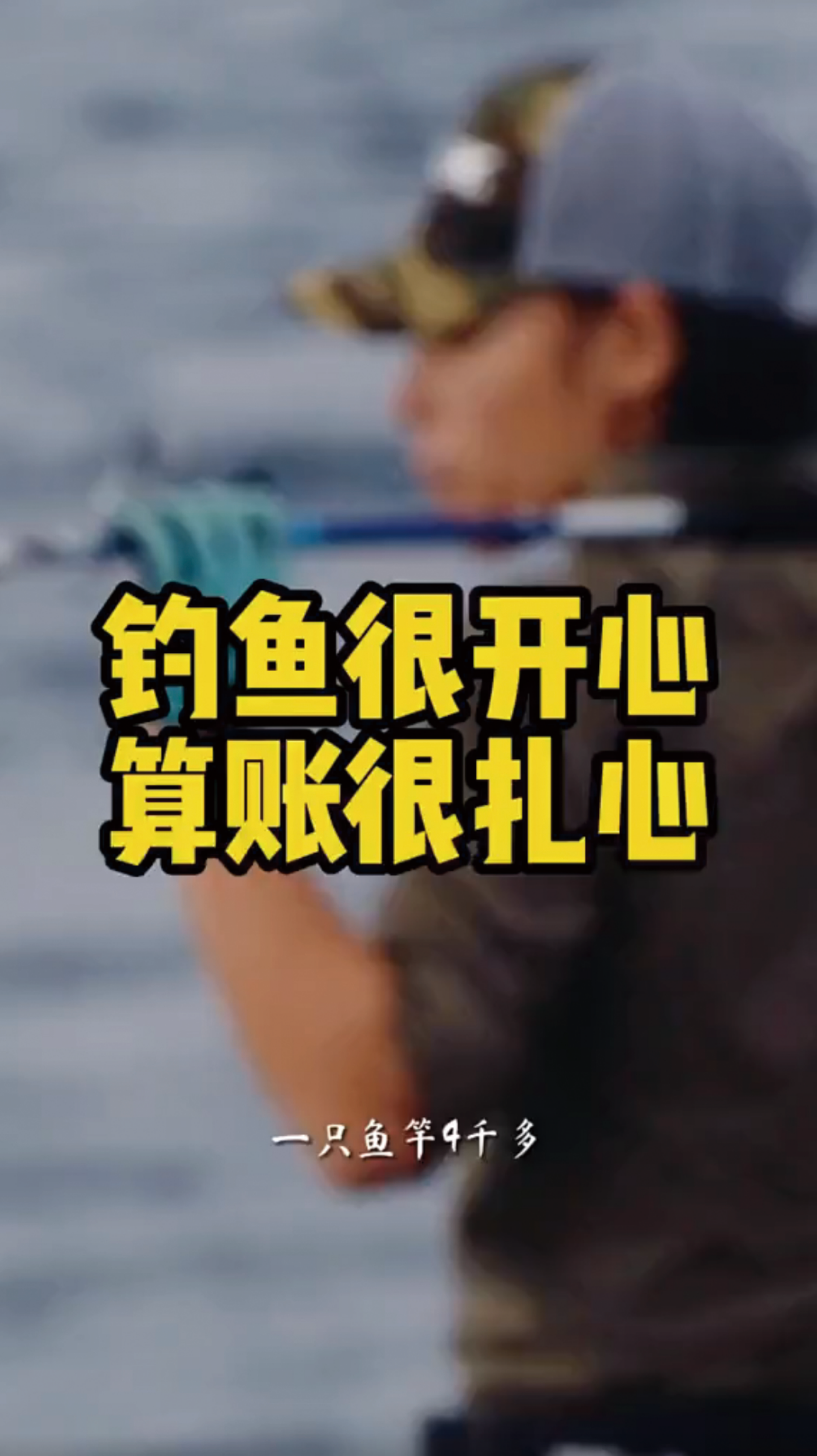}\end{minipage} & \begin{minipage}{0.06\textwidth}\includegraphics[width=1.1\linewidth]{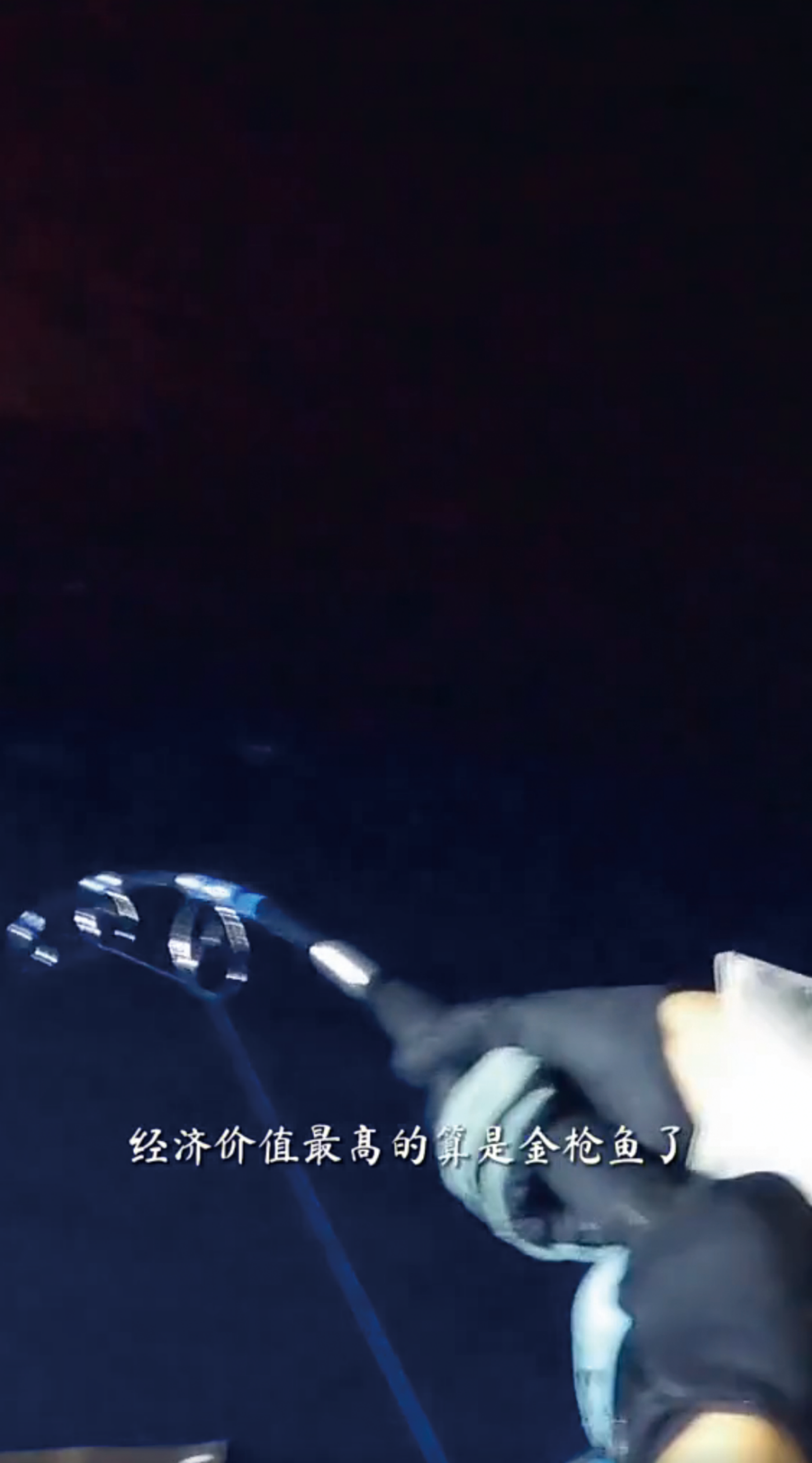}\end{minipage} & \begin{CJK*}{UTF8}{gbsn}\makecell{快来钓鱼,\\快手野钓月,\\哈利路亚,\\快手钓鱼}\end{CJK*} & \begin{CJK*}{UTF8}{gbsn}\makecell{休闲钓鱼,\\钓鱼心得,\\钓鱼乐趣,\\钓鱼爱好者,\\生活秀}\end{CJK*} & \begin{CJK*}{UTF8}{gbsn}\makecell{钓鱼乐趣,\\钓鱼放松,\\娱乐休闲,\\渔具装备,\\钓技攻略}\end{CJK*} & \begin{CJK*}{UTF8}{gbsn}\makecell{文本,\\运动,\\水,\\海报,\\钓鱼}\end{CJK*} & \begin{CJK*}{UTF8}{gbsn}\makecell{钓鱼,\\渔获,\\算账,\\快手}\end{CJK*} \\
    & & & & & \\
    \begin{minipage}{0.06\textwidth}\includegraphics[width=1.1\linewidth]{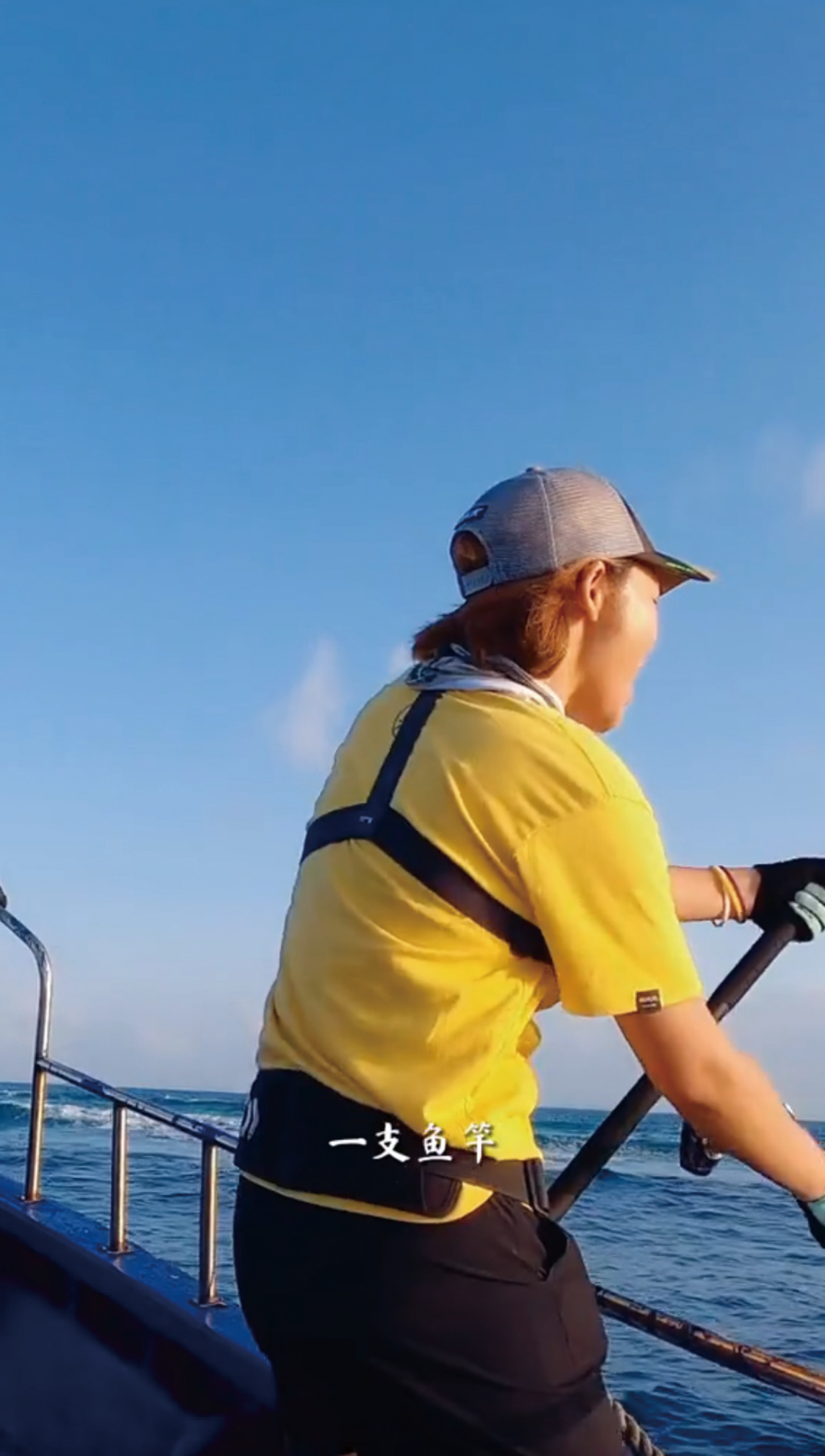}\end{minipage} & \begin{minipage}{0.06\textwidth}\includegraphics[width=1.1\linewidth]{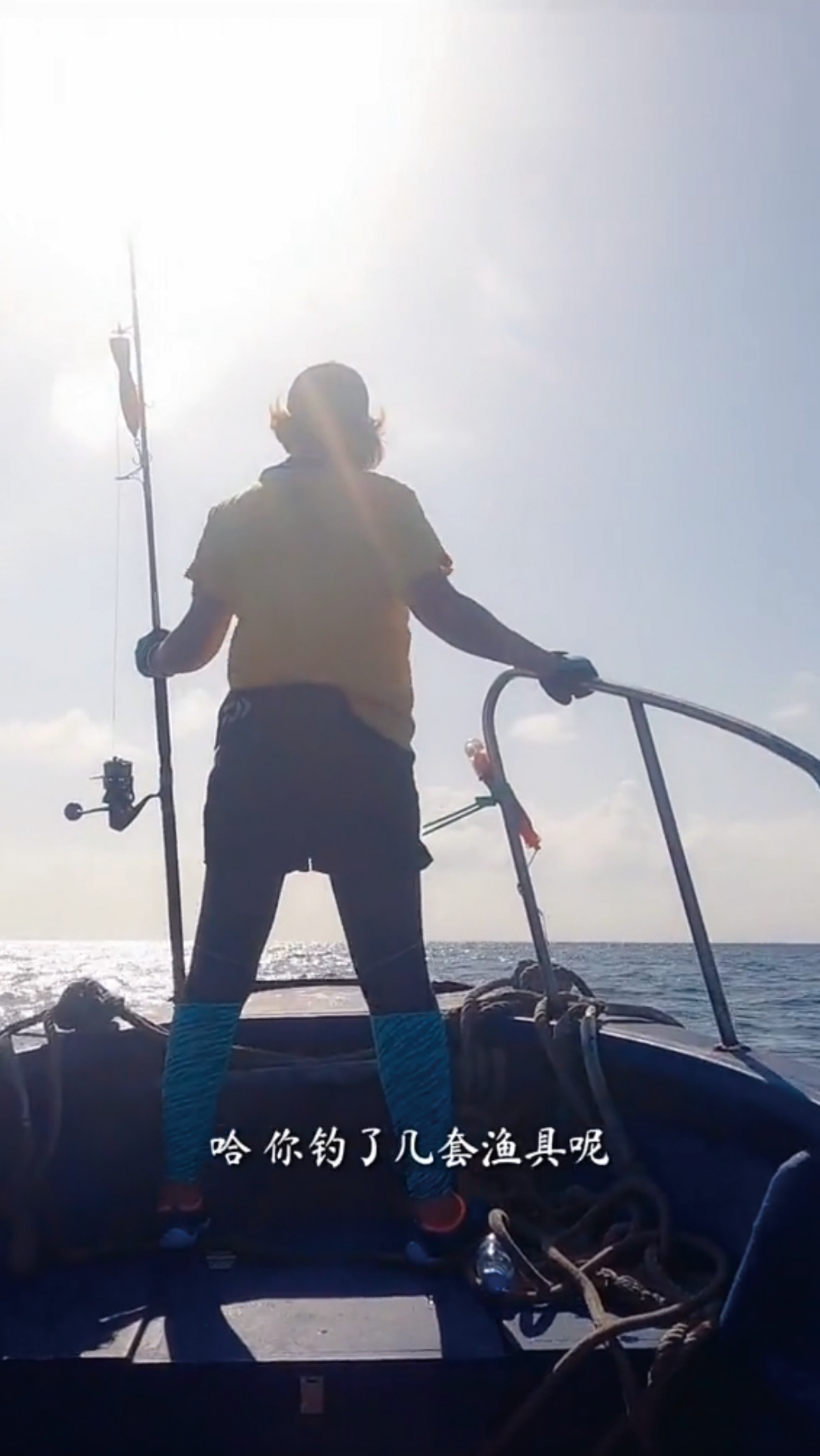}\end{minipage} & \makecell{come and fish,\\Kuaishou wild\\fishing month,\\Hallelujah,\\Kuaishou fishing} & \makecell{leisure fishing,\\fishing experience,\\fishing fun,\\fishing enthusiasts,\\life show} & \makecell{fishing fun,\\fishing relaxation,\\entertainment and leisure,\\fishing gear and equipment,\\fishing skills raiders} & \makecell{text,\\sports,\\water,\\poster,\\fishing} & \makecell{fishing,\\fishery harvesting,\\accounting,\\kuaishou} \\
    \midrule
    \begin{minipage}{0.06\textwidth}\includegraphics[width=1.1\linewidth]{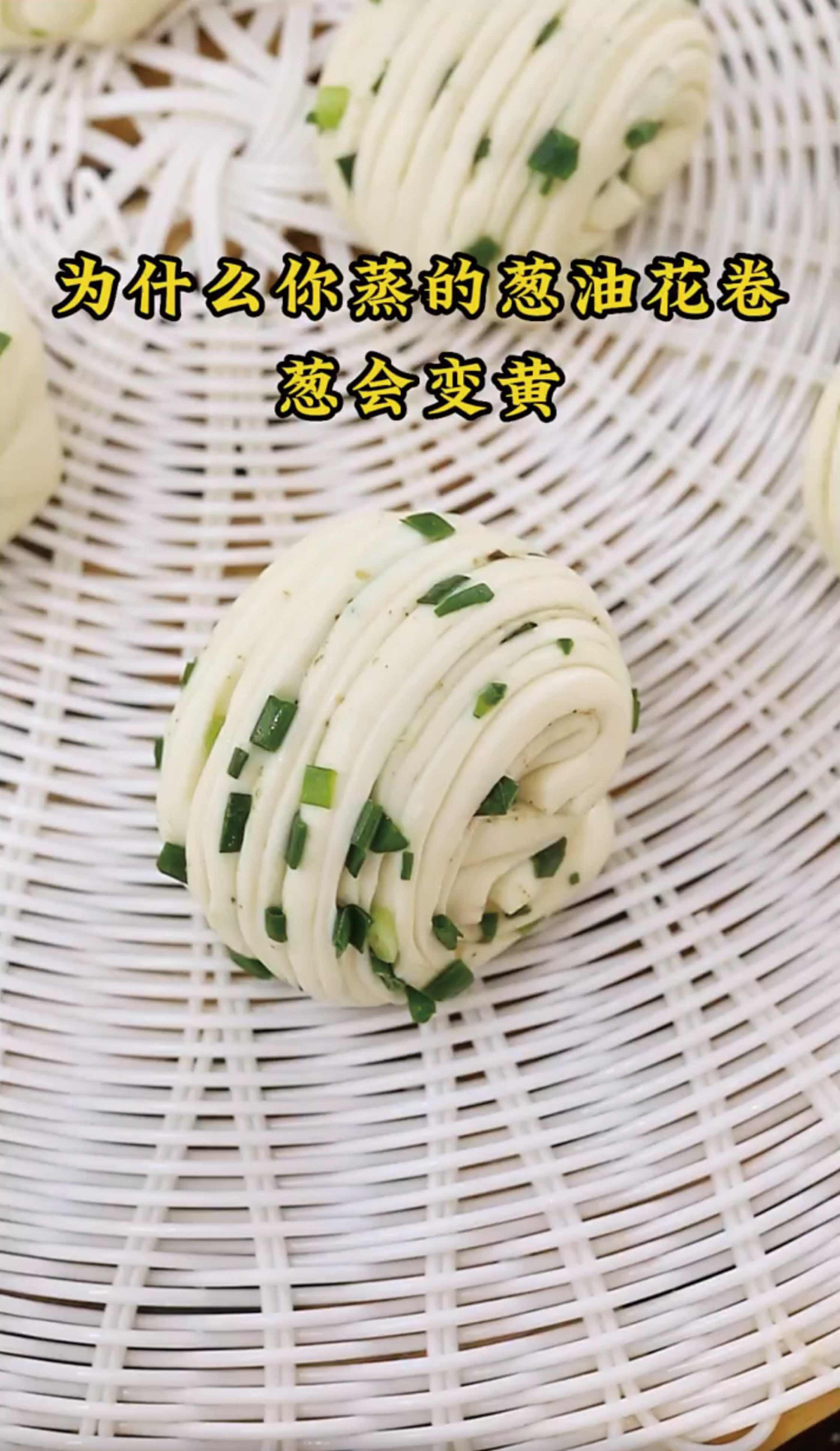}\end{minipage} & \begin{minipage}{0.06\textwidth}\includegraphics[width=1.1\linewidth]{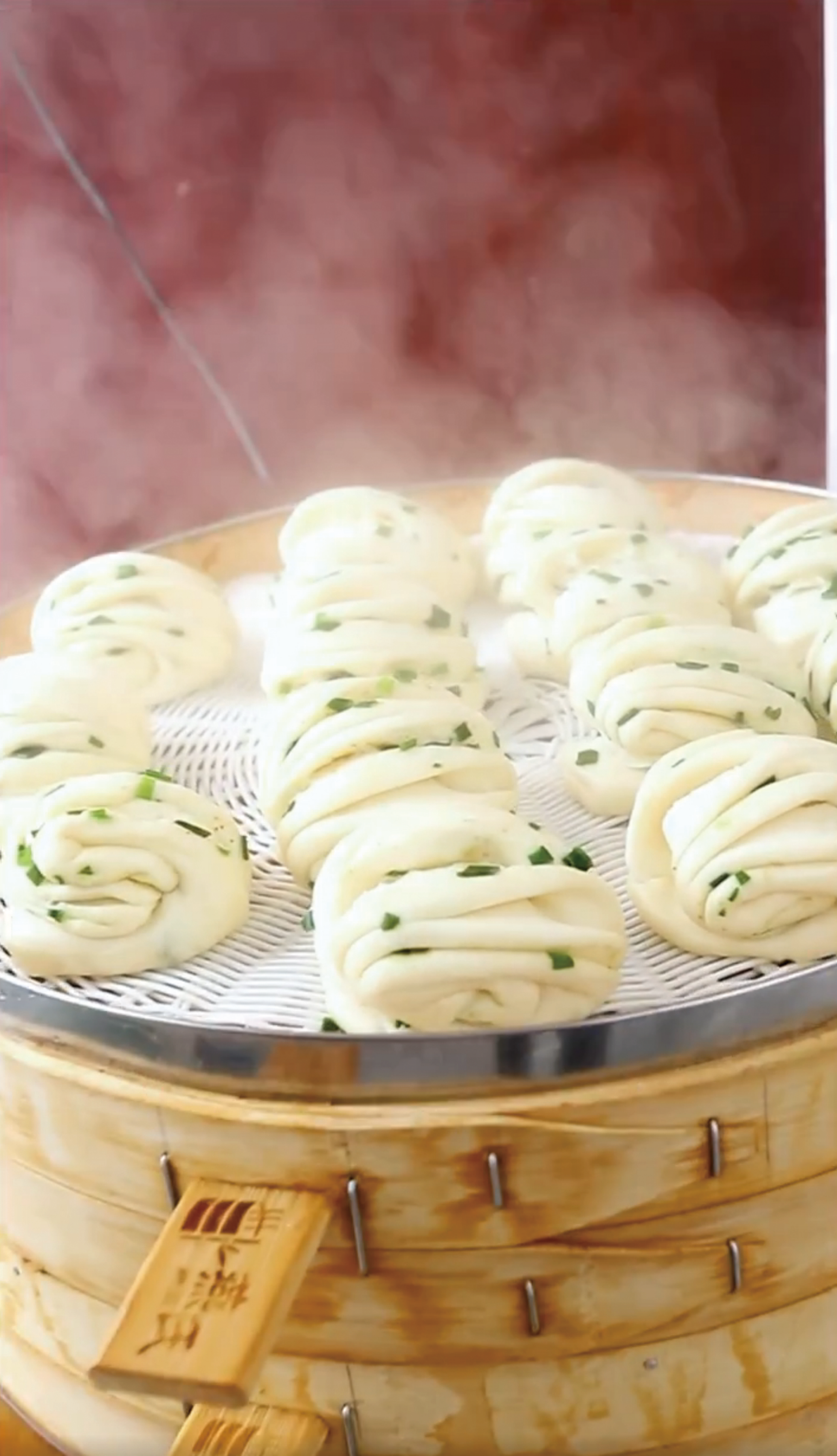}\end{minipage} & \begin{CJK*}{UTF8}{gbsn}\makecell{花卷,\\花样美食制作,\\面食}\end{CJK*} & \begin{CJK*}{UTF8}{gbsn}\makecell{花卷制作,\\自制面食技巧,\\美食掌故,\\家庭蒸菜,\\花样面食}\end{CJK*} & \begin{CJK*}{UTF8}{gbsn}\makecell{花卷制作,\\食用小方法,\\精选主食做法,\\葱油花卷制作技巧,\\早餐好吃的面食}\end{CJK*} & \begin{CJK*}{UTF8}{gbsn}\makecell{食品,\\蔬菜}\end{CJK*} & \begin{CJK*}{UTF8}{gbsn}\makecell{蒸葱,\\拌均匀,\\花卷,\\面食}\end{CJK*} \\
    & & & & & \\
    \begin{minipage}{0.06\textwidth}\includegraphics[width=1.1\linewidth]{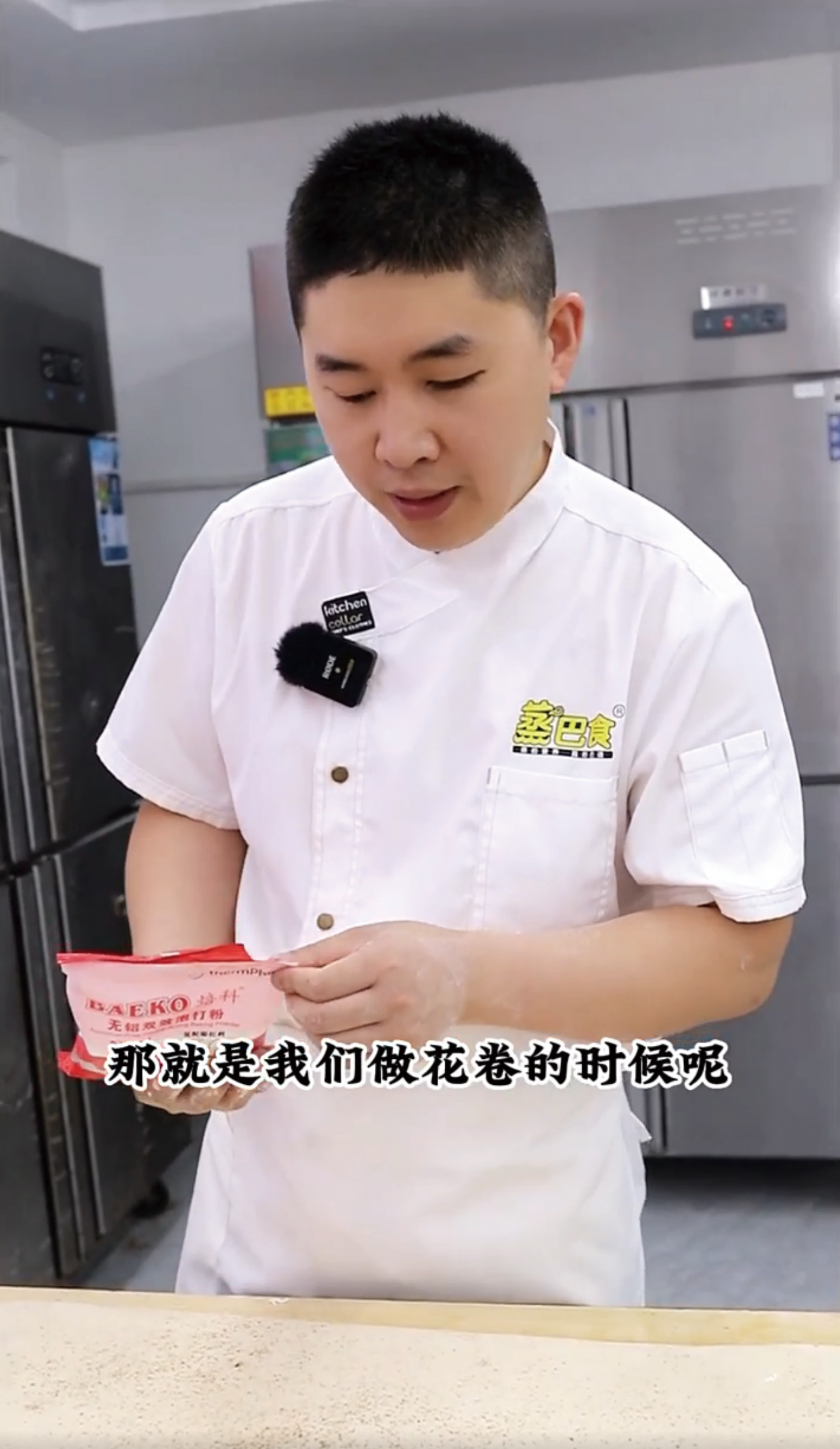}\end{minipage} & \begin{minipage}{0.06\textwidth}\includegraphics[width=1.1\linewidth]{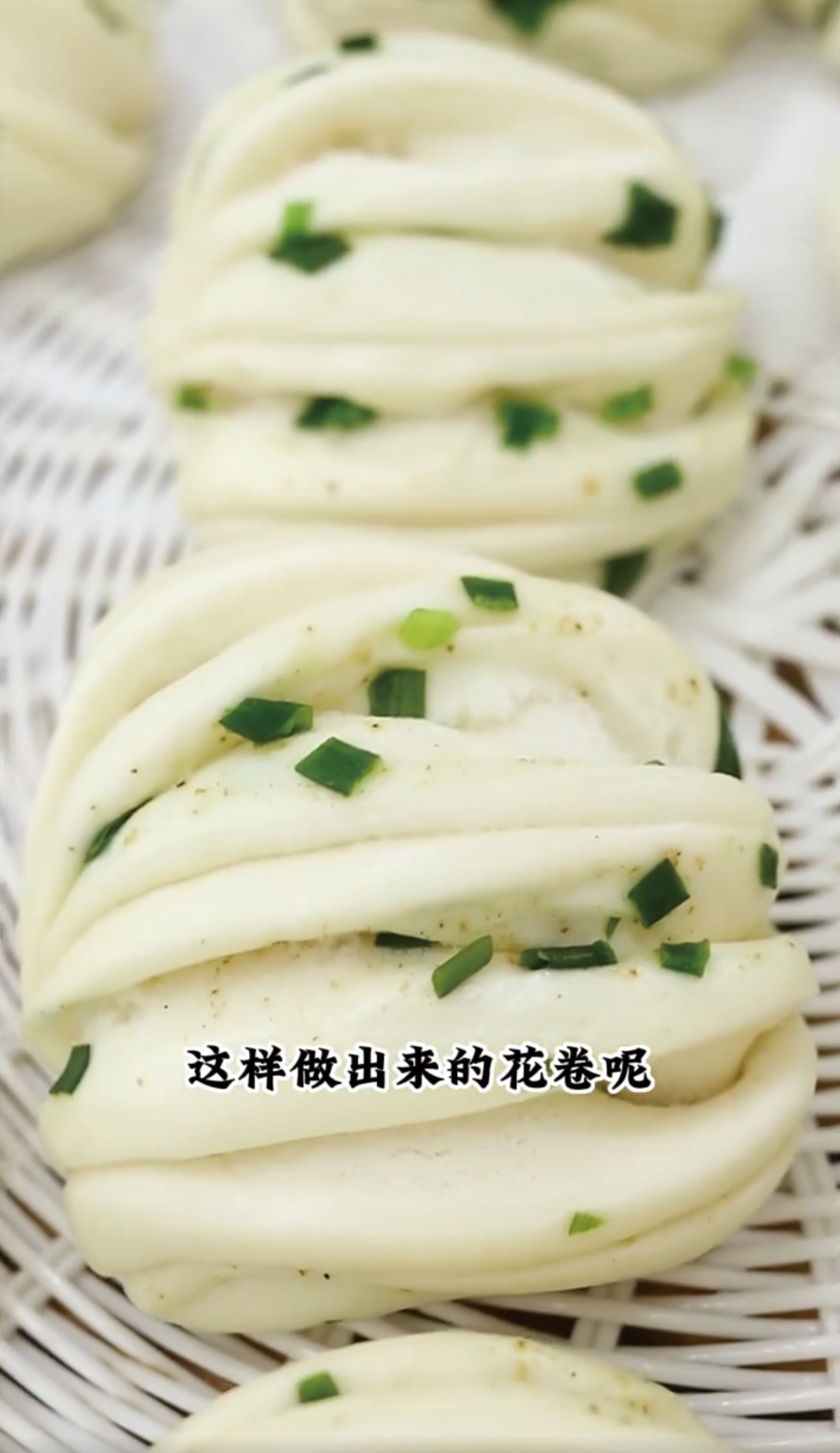}\end{minipage} & \makecell{steamed twisted\\roll, fancy\\food production,\\cooked wheaten\\food} & \makecell{steamed twisted roll production,\\homemade pasta skills,\\gourmet stories, \\home steamed dishes,\\pattern pasta} & \makecell{steamed twisted roll\\production, eating tips,\\selected staple food\\practices, scallion oil\\steamed twisted roll making\\skills, delicious pasta for breakfast} & \makecell{food,\\vegetables} & \makecell{Steamed shallots,\\stir evenly,\\steamed twisted roll,\\cooked wheaten food}\\
    \bottomrule
    \end{tabular}
    \vspace{0.1in}
    \caption{Qualitative results of \modelname. ACSV and Hanlp are \url{https://portal.vision.cognitive.azure.com/demo/generic-image-tagging} and \url{https://www.hanlp.com/demonstrate.html}.}
    \label{tab:case_studies}
\end{table}

A further comparison of the two different paradigm tagging models in \modelname reveals that both the selective tagger and the generative tagger can produce better tagging results than the human tagger. The selective tagger selectively assigns tags based on a subset of the roughly matched tags, leading to some degree of semantic coarseness while still being able to match the semantics of the given data. On the other hand, the generative tagger understands the given multimodal data and assigns tags in a more precise and accurate manner, fitting the content details more seamlessly. The selective tagger versus generative tagger comparison is essentially a trade-off between universality and precision in the tag assignment process. Overemphasizing universality can result in tags that are indistinguishable and lack analytical value, while overemphasizing precision can lead to the risk of over-interpretation, making it difficult to correctly associate relevant content.

%% file: texts/4.limitations.tex
\section{Limitations}\label{sec:limit}
It has been verified that \modelname is capable of automatically constructing a tagging system from large-scale multimodal data and providing tags for given data in a zero-shot setting. However, due to its design and technology, it is still subject to certain limitations:

\paragraph{Limitation of LLM}
Intuitively, the understanding and generation abilities of the LLM are directly related to the performance of \modelname. Therefore, when the performance of the LLM decreases, \modelname's performance will also be affected accordingly. Hence, it is recommended to utilize an LLM with robust performance or a language model that has undergone prompt tuning for optimal results.

\paragraph{Limitation of input length}
When the provided multimodal data contains excessive information, it can result in an excessively long input for the LLM, leading most LLMs to perform truncation operations that render the data unusable. This phenomenon renders \modelname ineffective in situations involving lengthy videos or texts.

\paragraph{Limitation of security and privacy}
As a lot of LLMs and unsupervised representation acquisition methods are derived from the API interface within the network, uploading a large amount of user and other data without caution may lead to significant security and privacy concerns regarding data leakage.

%% file: texts/5.conclusion.tex
\section{Conclusion}\label{sec:col}
We propose \modelname, a concise and low-coupling framework that provides a complete set of tools for tag system construction and tagging models. This framework enables users to easily obtain content tags of large-scale multimodal data and perform zero-shot tagging, which can serve various downstream applications such as search and recommendation.  To build a high-quality tag system based on understanding multimodal content, we introduce an LLM and an unsupervised text representation model. We also propose two different zero-shot tagging paradigms, both of which yield high-quality tagging results.  In the future, we aim to further optimize each sub-module in \modelname to improve its work efficiency and accuracy.

%% file: texts/6.appendix.tex
\appendix

\section{Instruction template}
\begin{table}[H]
    \centering
    \begin{tabular}{c|c}
        \toprule
        Dataset & Instruction template \\
        \midrule
        Kuaishou & \begin{CJK*}{UTF8}{gbsn}\parbox[c]{0.8\linewidth}{你是一个视频的兴趣标签生成机器人，根据输入的视频标题、类别、ocr、asr推理出合理的"兴趣标签"，以多个多于两字的标签形式进行表达，以顿号隔开。例如，给定一个视频，它的"标题"为"全屋嵌入式低音音响，主要是这个投影仪真的是爱了 "，"类别"为"房产家居"，"ocr"为"42平,一室一厅小户型"，"asr"为"看，远方灯火闪亮着光。你一人低头在路上。这城市越大，越让人心慌多向往，多漫长。祝一路行李太多伤。把最初笑容都淡忘。时光让我们变得脆弱，却坚强，让我在爱青青对你唱。我多想能多陪你唱。把什么生的风景对你讲。"，兴趣标签生成机器人推断出合理的"兴趣标签"为"小户型装修、一室一厅装修、装修效果图"。那么，给定一个新的视频，它的"标题"为"\{title\}"，"类别"为"\{category\}"，"ocr"为"\{ocr\}"，"asr"为"\{asr\}"，请推断出该视频的"兴趣标签"：}\end{CJK*} \\
        \midrule
        Food.com & \parbox[c]{0.8\linewidth}{You are a tag generation bot for a recipe video, which infers reasonable recipe tags based on the input video "dish name", "description" and "asr", and expresses it in the form of tags with more than two words, separated by commas. For example, given a recipe video whose "dish name" is "Grilled Garlic Cheese Grits", "description" is "We love grits, this is another good way to serve them. A great alternative to a baked potato when served with grilled steak or chicken. I believe this recipe could be made with instant grits. The 2 1/2 hours for refrigeration is not included in the time. The recipe comes from Taste of Homes Light and Tasty.", and "asr" is "In a saucepan, bring water to a boil; slowly add grits and salt, stirring constantly; Reduce heat: simmer, uncovered, for 40-45 minutes or until thickened, stirring occasionally. Add cheese and garlic; stir until cheese is melted, Spray a 9-inch baking dish with nonstick cooking spray;   Cover and refrigerate for 2 to 2 1/2 hours or until firm. Before starting the grill, coat the grill rack with nonstick cooking spray; Cut the grits into 3-inch squares; Brush both sides with olive oil. Grill, covered, over medium heat for 4 to 6 minutes on each side or until lightly browned.", the tag generation bot infers a plausible recipe tag as "time-to-make, course, main-ingredient, preparation, occasion, side-dishes, eggs-dairy, refrigerator, diabetic, vegetarian, grains, cheese, stove-top, dietary, low-cholesterol, low-calorie, comfort-food, low-carb, low-in-something, pasta-rice-and-grains, brunch, taste-mood, equipment, presentation, served-hot, 4-hours-or-less". Then, given a new video whose "dish name" is "\{name\}", "description" is "\{caption\}", and "asr" is "\{asr\}", please infer that the videos recipe tag is:} \\
        \bottomrule
    \end{tabular}
    \vspace{0.1in}
    \caption{Instruction template for tag generation in \modelname.}
    \label{tab:instruction_template}
\end{table}

In this section, we show the instruction template used for different languages and different datasets as shown in Table~\ref{tab:instruction_template}.